\documentclass[11pt,a4paper]{article}
\usepackage{jheppub}
\usepackage[utf8]{inputenc}

\author{Christian Copetti and}
\author{Jorge Fern\'andez-Pend\'as}

\affiliation{Instituto de F\'isica Te\'orica UAM/CSIC, c/Nicol\'as Cabrera 13-15, Universidad Aut\'onoma de Madrid, Cantoblanco, 28049 Madrid, Spain}

\emailAdd{christian.copetti@uam.es}
\emailAdd{j.fernandez.pendas@csic.es}

\title{Membrane paradigm and RG flows for anomalous holographic theories}

\abstract{
Holographic RG flows can be better understood with the help of radially conserved charges. It was shown by various authors that the bulk gauge and diffeomorphism symmetries lead to the conservation of the zero mode of the holographic $U(1)$ current and, if the spacetime is stationary, to that of the holographic heat current. In describing dual theories with 't Hooft anomalies the bulk gauge invariance is broken by Chern-Simons terms. We show that conservation laws can still be derived and used to characterize the anomalous transport in terms of membrane currents at the horizon. We devote particular attention to systems with gravitational anomalies. These are known to be problematic due to their higher derivative content. We show that this feature alters the construction of the membrane currents in a way which is deeply tied with the anomalous gravitational transport.
}

\begin{document}

\maketitle

\flushbottom

\section{Introduction}
During the last years there has been a renewed interest in the dynamics of gravitational theories on null horizons. The first works on this date back to the early days of the membrane paradigm (see \cite{Thorne:1986iy} for a review) and the idea that geometric fluctuations of the null geometry can be seen as hydrodynamic modes through the projection of Einstein's equations.

This construction, originally conceived to describe the dynamics of black holes in an isolated way, has seen a new interpretation in the light of holography. In fact, various works \cite{Heemskerk:2010hk,Faulkner:2010jy,Sin:2011yh} have connected the radial slicing of an asymptotically AdS spacetime with the Wilsonian RG scale of the dual theory. Although the relationship between the two quantities cannot in general be made explicit apart from very special examples, this interpretation offers the hint that, in a universal way, long wavelength fluctuations of the horizon geometry may be linked to low energy observables in the dual theory. In this sense, the universality of the Wilsonian low energy description is somewhat reflected in the general form of the horizon membrane equations. 

In order to make this comparison quantitative, however, one has to match the horizon data with the dual one point functions, which are given by specific limits of the bulk data once the conformal boundary is approached. In general this requires the full solution of the bulk dynamical equations, which can be conveniently expanded in a fluid-gravity approximation \cite{Hubeny:2011hd}.

The situation is however much simpler in some special cases, as was noted originally in \cite{Iqbal:2008by}. Here the universality of some DC properties in the dual theory, such as the conductivity and the shear viscosity to entropy ratio, were linked to the presence of conserved charges in the gravitational description. Interestingly, once computed on the conformal boundary, said charges coincide with the prescription for the dual field theory's currents.

This insight, which in some sense is a restatement of Gauss law, was then extended by other authors, notably \cite{Donos:2014cya, Donos:2015gia, Donos:2017oym} by analyzing the full gravitational dynamics in stationary spacetimes. 
In this works, apart from the conservation of the $U(1)$ current of \citep{Iqbal:2008by}, it is shown that the dual heat current
\footnote{In standard quantum field theory the heat current is the Noether current associated to the symmetry generated by a Killing field $\xi^a$, namely
\begin{equation}
Q^a= T^{ab} \xi_b + A_b \xi^b J^a \, .
\end{equation}
Its conservation follows from the conservation of the stress tensor and $U(1)$ current
\begin{equation}
D_a Q^a =\left( D_a T^{ab}  - F^{ba} J_a + A^b D_a J^a \right) \xi_b = 0 \, ,
\end{equation}
by using the Killing condition on all of the fields. Notice, however, that this is not a gauge independent quantity, but one can remedy this situation by adding an axion field $\theta$ which cancels the gauge variation
\begin{equation}
\tilde{Q}^a = T^{ab} \xi_b + \left(A_b - \partial_b \theta \right) \xi^b  J^a \, ,
\end{equation}
and the Killing condition on the Lie derivative becomes $\mathcal{L}_\xi A = d i_\xi d \theta$.
}
\begin{equation}
Q^i =\int_{\partial AdS}\left( J_\epsilon^i + A_t J^i \right)\, ,
\end{equation} 
where $J_\epsilon^i={T^i}_t$ is the energy current, is also radially conserved under these hypotheses. This makes it possible to link not only charge, but also energy fluctuations on the horizon to the infrared physics in the dual description.

Although their analysis doesn't make an explicit use of it, the form of the derived charge densities is closely related to the well known Komar charges of general relativity, which one can rigorously define by the Wald procedure \cite{Wald:1999wa}. This connection was pointed out for example in \cite{Liu:2017kml}. In fact, when the spacetime geometry admits a Killing vector, one is able to use the Wald construction to define a closed $d-1$ form $k$ which does not vanish on-shell. Then, as $\partial_\mu k^{\mu\nu}=0$, one can use the fluxes
\begin{equation}
F^\mu= \int_\Sigma d^d x \sqrt{-g} n_\nu k^{\mu\nu} 
\end{equation} 
to define conserved quantities with respect to the normal direction to $\Sigma$. Notice that here $d+1$ is the number of bulk dimensions. Various works have been applying this philosophy to various gravitational theories in order to analytically compute their DC properties during the last two years.

\bigskip

\noindent At the same time, but in a very different setting, holographic techniques were successfully used to model the low energy effects related to the presence of 't Hooft anomalies (see \cite{Kharzeev:2013ffa, Landsteiner:2016led} for recent reviews) in the dual theory by supplementing the bulk action with the appropriate Chern-Simons term \cite{Erdmenger:2008rm, Banerjee:2008th}.

In particular, the interest was focused on four dimensional fermionic theories with an anomalous chiral current
\begin{equation}
D_\mu J^\mu_5 = -\kappa \epsilon^{\mu\nu\rho\sigma} F_{\mu\nu} F_{\rho\sigma} -\lambda \epsilon^{\mu\nu\rho\sigma} {\rm tr} \left(\bf{R}_{\mu\nu} {R}_{\rho\sigma}\right) \, . \label{anomalyinintroduction}
\end{equation}
For this theory it was known that the currents would develop an anomalous DC response of the form
\begin{align}
J_5^\mu&= \sigma_B B^\mu + \sigma_\omega \omega^\mu + \ldots \, , \\
J^\mu_\epsilon&= \xi_B B^\mu + \xi_\omega \omega^\mu + \ldots \, ,
\end{align}
where the magnetic field and the vorticity are, respectively, $B^\mu= \frac{1}{2}\epsilon^{\mu\nu\rho\sigma}u_\nu F_{\rho\sigma}$ and $\omega^\mu= \frac{1}{2}\epsilon^{\mu\nu\rho\sigma}u_\nu \partial_\rho u_\sigma$, and the ellipses stand for the dissipative and ideal parts of the fluid expansion.
\footnote{These values depend on the frame choice. In this work we are however going to compute the one-point functions in a particular choice of coordinate system that recovers the values above.}
In particular it was shown by \cite{Son:2009tf} that most of the unknown coefficients above are uniquely fixed by demanding consistence of the hydrodynamic expansion in terms of the chiral anomaly coefficient $\kappa$
\begin{equation}
\sigma_B= -8 \kappa \mu \, , \ \ \sigma_\omega= 2 \xi_B= -\left( 8 \kappa \mu^2 + \gamma T^2  \right)  \, , \ \ \xi_\omega=  -\left( \frac{16}{3} \kappa \mu^3 + 2 \mu \gamma T^2 \right) \, . \label{Ianomaloscoefficients}
\end{equation}
The undetermined coefficient $\gamma$ was linked in \cite{Landsteiner:2011cp} to the mixed gauge gravitational anomaly for massless fermions
\begin{equation}\label{eq:cvccoeff}
\gamma= 64  \pi^2 \lambda \, ,
\end{equation}
and confirmed at strong coupling by holographic methods \cite{Landsteiner:2011iq}. The universality of this contribution in the hydrodynamic expansion is somewhat surprising, as the gravitational anomaly itself only enters at third order in derivatives. 

Various arguments have since been developed in order to fix this coefficient by equilibrium considerations only \cite{Jensen:2012kj, Jensen:2013rga, Jensen:2013kka}. In the context of holography the emergence of this term in the near horizon fluctuations of the $U(1)$ constraints was shown in \cite{Chapman:2012my}. Its link to the dual one point functions was made explicit in \cite{Azeyanagi:2013xea}, where it was shown how to recast these observables in term of horizon quantities by explicitly solving the bulk equations of motion. Furthermore it was recently showed by various authors \citep{Golkar:2015oxw, Glorioso:2017lcn} that a part of this contribution can be explained through the matching of the \emph{global} gravitational anomalies.
 
\bigskip

\noindent In this work we connect the aforementioned two lines of research, i.e. holographic conserved charges and anomalous transport. In order to do so, we show how the membrane paradigm can be extended to bulk Chern-Simons theories by suitably modifying the definitions of the membrane currents and stress tensor. To this end, we will need to examine the constraint equations on arbitrary hypersurfaces of constant radial coordinate to properly define the membrane observables and show that these match with the conserved quantities given by the Komar charges.

The Wald construction is slightly technical in this case due to the noncovariant nature of the Chern-Simons action for the bulk theory. It was generalized to these actions in various works \cite{Tachikawa:2006sz, Bonora:2011gz, Azeyanagi:2014sna}.
Other authors have used this formalism to perform a near horizon analysis of the entropy current  \cite{Chapman:2012my}. An argument closer in spirit to the work of Iqbal and Liu was given by \cite{Gursoy:2014boa, Grozdanov:2016ala}, where the conserved quantity is constructed for the $U(1)$ currents in an anomalous theory. They however lack an explicit argument for energy fluctuations.

In this note we aim to generalize and unify the various insights above. In particular, we include the energy current transport and we explicitly relate it to the membrane paradigm by the construction of the membrane currents for the anomalous theory. This last construction is not trivial, as bulk gravitational Chern-Simons terms give rise to higher derivative theories for which the Cauchy problem is in general not well defined. This gives rise to subtleties in defining the membrane stress tensor away from the conformal boundary. Furthermore, it is well known that in anomalous theories the current operators have no unique definition but one can define different operators depending on the properties which are desired e.g. consistent and covariant currents. We shed light on the role of these different operators from the point of view of the bulk dynamics.

We will then use the appropriate conserved charges to match the membrane results to the boundary observables, highlighting the dynamical mechanisms that give rise to the anomalous contributions from our perspective. In particular, we will show that the gravitational contributions to anomalous transport come from horizon extrinsic Chern-Simons currents which are dynamically generated along the radial direction. 

\bigskip

\noindent The work is organized as follows: in section 2 we review the extraction of the Wald charges in Einstein-Maxwell theory, discussing their link to the bulk constraint equations, and the construction of the membrane current and stress-tensor.
In section 3 we use the constraint equations of the bulk theory to propose a definition of membrane current and stress-tensor. We give some consistency checks for this proposal. In section 4 we review the generalized Wald construction as done by \cite{Bonora:2011gz, Azeyanagi:2014sna} and show that the continuity equations for the resulting close $d-1$ forms can be understood as RG equations for the previously defined membrane currents. We integrate them to recover the known anomalous transport coefficients and interpret them in the spirit of membrane paradigm. 

As the topic of anomalous transport in holography has already been widely studied let us add a few comments on the differences and similarities with previous works:
\begin{itemize}
\item As it was mentioned, the general conservation equations for the $U(1)$ currents were already derived by \cite{Gursoy:2014boa, Grozdanov:2016ala} by direct analysis of the equations of motion. They used this to prove the universality of the $U(1)$ transport. With respect to these works we extend the discussion to energy fluctuations and formalize the connection between the conserved quantities and the anomalous Wald construction. Historically, such a structure was hinted by the work of \cite{Azeyanagi:2013xea}, where the value of the renormalizable mode of the bulk fields is shown to be given by an horizon quantity through the explicit solution of the dynamical equations. 
\item The connection between the conserved fluxes of \cite{Donos:2015gia} and the Wald construction is not new, but has, to our knowledge, been formalized only recently \cite{Liu:2017kml}, while this work was being completed. Its application to the anomalous theories, however, had yet to appear. We formalize such extension and identify the conserved fluxes as the suitably defined membrane currents of the theory.
\item The constraint equations were already analyzed in the near horizon region in \citep{Chapman:2012my} which however lack an explicit link to the membrane currents.
\end{itemize}

Lastly, some comments about notations and conventions. We will be working in many cases through an ADM expansion of the bulk spacetime. In order to define the $d+1$ decomposition we will fix the Fefferman-Graham gauge for the bulk metric
\begin{equation}
ds^2 = dr^2 + \gamma_{ab} dx^a dx^b \, ,
\end{equation}
and foliate the spacetime with surfaces $\Sigma$ of constant $r$. We will work with spacetimes which are asymptotically Anti de Sitter, so that the following boundary conditions are imposed on the metric coefficients near $r=\infty$
\begin{equation}
\gamma_{ab} \sim e^{2r}\left( \hat{\gamma}_{ab} + O(e^{-r}) \right) \, . 
\end{equation}
$\hat{\gamma}$ is to be interpreted as the curved background metric of the CFT. This particular foliation has a nice interpretation in holography, as transitions from one $\Sigma$ to another is asymptotically viewed as a scale transformation and can thus be related to the Wilsonian flow of the dual theory.

We will use Greek indexes $\mu,\nu,...$ to denote bulk tensors, while tensors on $\Sigma$ are denoted with Latin letters $a,b,...$. When confusion may arise, quantities intrinsic to $\Sigma$ are denoted by hatted names $\hat{R}_{abcd}$ etc. The covariant derivatives are finally denoted by $\nabla_\mu$ and $D_a$ respectively.

In many cases we will not give the details of the ADM decomposition of all the quantities, but the results can be recovered through the following dictionary
\begin{align}
- \Gamma_{ab}^r &= K_{ab} \, , \\
\Gamma_{ar}^b &= K_a^b \, .
\end{align}

\section{Membrane currents and conserved charges for Einstein-Maxwell theory}

In this section we present a simple example in order to familiarize the reader with the main constructions and methods. We consider Einstein-Maxwell theory with cosmological constant in $d+1$ spacetime dimensions, whose action is given by
\begin{equation}\label{eq:EinsteinHibertaction}
S_g= \frac{1}{16 \pi G}\int d^{d+1}x \sqrt{-g}\left( R - 2 \Lambda\right) - \frac{1}{4}\int d^{d+1}x \sqrt{-g} F_{\mu\nu} F^{\mu\nu} + S_{GH} \, ,
\end{equation}
where
\begin{equation}
S_{GH}= \frac{1}{8\pi G} \int_\Sigma d^d x \sqrt{-\gamma} K \, ,
\end{equation}
is the Gibbons-Hawking counterterm, which assures a well defined variational problem at a general Cauchy surface. The considerations presented remain however valid, \emph{mutatis mutandis} for covariant action with no charged matter and we will try to use a general notation to make such generalizations manifest. In what follows we will employ an on-shell formalism, and denote equality up to equations of motion by the $\doteq$ symbol. 

\subsection{Membrane currents}
We will start by giving a procedure to define membrane currents and stress tensor. From our point of view they are fields living on a spacelike hypersurface $\Sigma$ which are asked to satisfy the usual field theory Ward identities.
By using the constraint equations of the bulk theory one can define such fields as functions of bulk quantities projected on $\Sigma$. As it is well known, these currents will in general need holographic counterterms to be finite. However, this inclusion does not spoil the form of the Ward identities and will not be relevant for the DC properties we compute.

The bulk constraint equations reduce to the field theory Ward identities on the conformal boundary. This suggests that, in the rest of the bulk, the constraint equations should be interpreted as Ward identities in the effective theory description at low energies.

To derive the form of the constraints, we need to cut the integration in \eqref{eq:EinsteinHibertaction} at an hypersurface $\Sigma$ of constant radial coordinate. The on-shell variation of the action reads
\begin{equation}\label{eq:variationEH}
\delta S_\Sigma = \int_\Sigma d^d x \sqrt{-\gamma}\left( \frac{1}{2} t^{ab} \delta \gamma_{ab} + J^a \delta A_a\right) \, , 
\end{equation}
where 
\begin{align}
t^{ab} &= -\frac{1}{8\pi G}\left( K^{ab} -\gamma^{ab} K \right) \, , \label{IIBYork} \\ 
J^a &= 2 n_\mu \frac{\partial L}{\partial F_{\mu\nu}} P^a_\nu = - n_\mu F^{\mu\nu} P^a_\nu \, ,  \label{IInormalcurrent} 
\end{align} 
are the Brown-York tensor and the membrane current, respectively, and $\gamma$ is the induced metric on $\Sigma$. We also have defined the vector $n^\mu$, normal to $\Sigma$, and the orthogonal projector $P_\mu^a$ onto $\Sigma$. In the Fefferman-Graham gauge, however, these are simply $n=\partial_r$ and $P_\mu^a=\delta_\mu^a$, so that we will sometimes omit them for ease of notation.

One can then write, as customary
\begin{align}
t^{ab} & \doteq\frac{2}{\sqrt{-\gamma}} \frac{\delta S_g}{\delta \gamma_{ab}} \, , \\
J^a & \doteq  \frac{1}{\sqrt{-\gamma}}\frac{\delta S_g}{\delta A_a} \, .
\end{align}

The constraints are then found by considering the gauge and diffeomorphism variation in \eqref{eq:variationEH}. They read
\begin{align}\label{eq:standardWard}
D_a J^a &= 0 \, , \\ 
D_a t^{ab} - F^{ba} J_a &= 0 \, .
\end{align}
These are exactly the Ward identities for the $U(1)$ current and the stress tensor in standard field theory.

These currents may be defined at every $\Sigma$, but they are in general non trivially related between one surface and the other by explicitly solving the dynamical equations, which can be expressed as first order PDE for the membrane observables.

\subsection{Wald charges and conservation laws}

As we have anticipated in the introduction, there are conserved charges that can be used to relate the zero modes, or fluxes, of these currents between different membranes. They can be derived through the Wald construction as follows.

Start with the variation of the bulk action
\begin{equation}
\delta S_\Sigma = \int E \delta \Phi + d \theta \, , \label{variationofaction}
\end{equation}
where $E$ denotes the equations of motion and $\Phi$ the set of fields. As a matter of notation the $d$-form $\theta= \theta(\Phi, \delta \Phi)$ is called the presymplectic form.

Now suppose that the variation is made with respect to a diffemorphism generated by $\xi$ and a gauge variation generated by $\alpha$. Then, $\delta = \delta_\xi + \delta_\alpha$. Supposing the action to be covariant, i.e. no anomalies present, the equation \eqref{variationofaction} can be rewritten using Cartan's formula to give the on-shell closure relation
\begin{equation}
d J_{\xi, \alpha} \doteq 0 \, \,
\end{equation}
for the Noether current
\begin{equation}
J_{\xi, \alpha} \doteq \theta_{\xi, \alpha} - i_\xi L \, , \label{eq:onshellvanishingcurrent}
\end{equation}
where $L$ is the Lagrangian. This current has various ambiguities, as pointed out by Wald \cite{Wald:1999wa}. In particular, we can add to it an exact form $dk$ without spoiling its closedness 
\begin{equation}
 J_{\xi,\alpha} \rightarrow \hat{J}_{\xi,\alpha} = \theta_{\xi,\alpha} - i_\xi L + d k_{\xi,\alpha} \, .
\end{equation}
We fix this ambiguity by demanding that 
\begin{equation}
\hat{J}_{\xi, \alpha} \doteq 0 \, . 
\end{equation} 
This is always possible as $\hat{J}$ is locally exact on-shell. The vanishing of the Noether current implies a closure relation for $k$ when the gauge transformation or diffemorphism leaves the background solution invariant. We briefly review how this is done in the relevant cases.

Let us start by considering a pure gauge transformation that leaves the field configuration invariant. As $\delta_\alpha A = d \alpha$, this forces $\alpha=\rm{const}$. For such transformations the presymplectic form vanishes identically, as it is proportional to $d \alpha$. Then, plugging this in \eqref{eq:onshellvanishingcurrent} we get our desired closure relation
\begin{equation}
0 \doteq \hat{J}_\alpha \doteq d k_\alpha \, , \hspace{0.5cm} {\rm for} \ \alpha = \rm{const} \, .
\end{equation}
The closed form $k$ can be integrated on a $d-1$ dimensional hypersurface and one recovers the Gauss law. However, in holography we will be interested in its fluxes over $d$ dimensional surfaces $\Sigma$ of constant radial coordinate. 
These are defined in terms of its Hodge dual $k^{\mu\nu}$ as
\begin{equation}
I^a \doteq \int_\Sigma d^d x \sqrt{-g} \left( n_\mu k^{\mu\nu}_\alpha P_\nu^a \right) \, .
\end{equation}
The closure relation $\partial_\mu k^{\mu\nu} \doteq 0$ then translates to the radial conservation of the current fluxes
\begin{equation}
\partial_r I^a \doteq -\int_{\Sigma} d^d x \partial_b \left( \sqrt{-\gamma} k^{ab}_\alpha \right) = 0 \, ,
\end{equation}
as long as the surfaces terms go to zero sufficiently fast. This is not the case when some DC modes are present. Then, extra care is needed.

In the context of holography this is to be interpreted as an RG equation for the conserved quantities $I^a$, meaning that they can be exactly extracted from the dynamics of fluctuations very far away form the conformal boundary.

In the case of diffeomorphisms the natural generalization is to impose a Killing condition on the set of bulk fields $\mathcal{L}_\xi \Phi \doteq 0$. This assures that the solution is left invariant by the diffeomorphism generated by $\xi$.\footnote{Notice that in the case of the gauge field a slightly more general condition is possible, namely $\mathcal{L}_\xi A = d \alpha_\xi $. The resulting charge is then given by a linear combination of the diffeomorphism and the gauge ones, by choosing the gauge parameter $\alpha= - \alpha_\xi$ in order to make the presymplectic form vanish.} 
Notice, however, that from the point of view of the solution, while the trivial gauge transformation $\alpha= \rm{const}$ is always present independently of the dynamics, the Killing equations are not satisfied for general backgrounds and place dynamical restrictions on the construction of the diffeomorphism charge.

The existence of a Killing vector assures, as before, that the presymplectic current vanishes on-shell, since it is by construction proportional to $\mathcal{L}_\xi \Phi$. The extra term proportional to the Lagrangian can always be written as a total derivative, as long as it is covariant. In fact, for covariant Lagrangians
\begin{equation}
0 \doteq \mathcal{L}_\xi L = d i_\xi L \, , 
\end{equation}
and, therefore,
\begin{equation}
i_\xi L \doteq d \zeta_\xi \, ,
\end{equation}  
by Poincare's lemma.
Then one can define a new quantity $k'_\xi = k_\xi -\zeta_\xi$ which is a closed $d-1$ form
\begin{equation}
0 \doteq \hat{J}_\xi \doteq d k'_\xi \, .
\end{equation}
In a similar way to the gauge case, one can define a diffeomorphism conserved flux $H^a$ by integrating $k'$ on a $d$ dimensional surface
\begin{equation}
H^a = \int_\Sigma d^d x \sqrt{-g} \left( n_\mu {k'_\xi}^{\mu\nu} P_\nu^a \right) \, ,
\end{equation}
which is radially conserved
\begin{equation}
\partial_r H^a \doteq 0 \, .
\end{equation}
In our application, it is customary to take the Killing vector $\xi$ to be associated with time translations, i.e. $\xi= \partial_t$. Furthermore, this result is usually presented as a spatial flux $H^i$. For this components the computation gets simplified, as the Lagrangian term in the Noether current gives no contributions and
\begin{equation}
H^i = \int_\Sigma d^d x \sqrt{-g} \left( n_\mu k_\xi^{\mu\nu} P_\nu^i \right) \, .
\end{equation}
However, in general the ``time'' component of the flux is also radially conserved.

\bigskip

\noindent The holographic importance of this construction becomes apparent once the conserved fluxes are written as functions of our previously defined membrane currents. We will use a direct construction. However, from a general standpoint the link between the membrane currents and the Komar charge should be encoded in the relation between the Wald construction and the constraint equations.

For Einstein-Maxwell theory and, in general, for actions depending on the gauge field only through the curvature, the presymplectic form associated to gauge transformations is given by
\begin{equation}
\theta_\alpha= \frac{\partial L}{\partial F} d \alpha \, ,
\end{equation}
or, in components,
\begin{equation}
\theta^\mu_\alpha= 2 \frac{\partial L}{\partial F_{\mu\nu}} \nabla_\nu \alpha \, .
\end{equation}
As in this case the equations of motion read $\nabla_\mu \frac{\partial L}{\partial F_{\mu\nu}} = 0$, we can easily make the Noether current vanish on-shell by adding $\nabla_\nu k^{\mu\nu}$,  where
\begin{equation}
k^{\mu\nu}= - 2 \frac{\partial L}{\partial F_{\mu\nu}} \alpha \, .
\end{equation}
The final Noether current is
\begin{equation}
\hat{J}^\mu_\alpha = - 2 \alpha \nabla_\nu \frac{\partial L}{\partial F_{\mu\nu}} \doteq 0 \, .
\end{equation}

Finally, confronting $n_\mu k^{\mu\nu}_\alpha P_\nu^a $ with \eqref{IInormalcurrent} one sees immediately that the normal-tangent components of $k^{\mu\nu}$ correspond to the membrane currents of the bulk hypersurfaces if one fixes the constant $\alpha=1$. The radially conserved flux $I^a$ is, as noted by \cite{Donos:2014uba}, just the flux of such currents
\begin{equation}
 I^a = \int_\Sigma d^d x \sqrt{-\gamma} J^a \, ,
\end{equation}
which, by construction, coincides with the flux of the dual CFT's $U(1)$ current. As mentioned before, this allows to match the horizon membrane paradigm to the dual theory's predictions.

For the diffeomorphism part a slightly more involved computation is needed. Let us start by recalling that, for the Einstein-Maxwell theory, the presymplectic current is given by \citep{Iyer:1994ys}\footnote{In general, for theories depending only on the curvatures ${{R_{\mu\nu}}^\rho}_\sigma$ and $F_{\mu\nu} $ one has
\begin{equation}
\theta^\mu = 2 \frac{\partial L}{\partial {{R_{\mu\nu}}^\rho_\sigma} } \delta {\Gamma_\nu}^\rho_\sigma + 2 \frac{\partial L}{\partial F_{\mu\nu}}  \delta A_\nu \, .
\end{equation}
}
\begin{equation}
\theta_{EH}^\mu=\frac{1}{16\pi G}g^{\mu\nu}g^{\rho\sigma}\left( \nabla_\rho \delta g_{\nu\sigma} - \nabla_\nu \delta g_{\rho \sigma} \right)  + 2\frac{\partial L}{\partial F_{\mu\nu}} \delta A_\nu \, , 
\end{equation}
which for a diffeomorphism becomes
\begin{equation}
\theta_{EH}^\mu= \frac{1}{16 \pi G} \nabla_\nu \left( \nabla^\nu \xi^\mu - \nabla^\mu \xi^\nu\right) + 2\frac{\partial L}{\partial F_{\mu\nu}} \left( \nabla_\nu (A_\rho \xi^\rho) + F_{\nu\rho} \xi^\rho \right) \, . \label{eq:thetapresymplEH}
\end{equation}
The Noether current can then be constructed by adding $\xi^\mu L$ to this term. To define the Komar form one just needs to add a total derivative to make the resulting expression vanish on-shell. A careful rewriting shows that the derivative of
\begin{equation}
k_{EH}^{\mu\nu} = \frac{1}{4\pi G} \nabla^{[\mu} \xi^{\nu]} + A_\alpha \xi^\alpha \frac{\partial L}{\partial F_{\mu\nu}} \, ,
\end{equation}
reduces the Noether current to a linear combination of Einstein and Maxwell equations. Following our previous construction one looks at the spatial flux density 
\begin{equation}
2 n_\mu k_{EH}^{\mu\nu} P_\nu^i = -\frac{1}{8\pi G} K^{i b} \xi_b + A_c \xi^c J^i \, ,
\end{equation}
where we have used that in Fefferman-Graham coordinates $\nabla_r \xi^a = -K^{ab} \xi_b$, if $\xi^a$ is radially independent.

One then recognizes the conserved flux associated to diffeomorphisms to coincide with the flux of the membrane heat current
\begin{equation}
H^i = \frac{1}{2} \int_\Sigma d^{d}x \sqrt{-\gamma}\left( {t^i}_b \xi^b + A_c \xi^c J^i \right)=\frac{1}{2} \int_\Sigma d^d x \sqrt{-\gamma} Q^i \, ,
\end{equation}
which, on the conformal boundary, tends to the dual theory's heat current flux, up to a factor of $1/2$ which is fixed by properly normalizing the diffeomorphism generator.

Let us conclude this section with two important remarks. First, notice that, although we have considered the Maxwell Lagrangian, the explicit dependence of the action on $F_{\mu\nu}$ never appears. Thus, the construction can be extended to all the Lagrangians which are functions of the curvature $F_{\mu\nu}$ and possibly to uncharged matter. For the gravity side the correspondence between fluxes and CFT observables was explicitly proven for various higher curvature theories, such as Gauss-Bonnet, $f(R)$ etc. and should hold as long as the bulk action is covariant and its Cauchy problem is well defined.

Second, the construction of the Komar charges is independent of the addition of possible surface counterterms to the action, as is customary in holographic renormalization. This is a consequence of the on-shell nature of the construction. A way of rephrasing this is by noticing that the divergence free currents $k^{\mu\nu}$ can be constructed directly from the Einstein and Maxwell equations of motion. In the gauge case the construction is apparent and was the basis for the seminal work \cite{Iqbal:2008by}. In the Einstein case one needs a bit more work and, in particular, one needs to consider the Einstein equations dotted with the Killing field $\xi^\mu$ and use various identities related to Lie derivatives, as in \citep{Donos:2015gia}.

As the dynamical equations are counterterm-independent this assures that, as long as the charges are finite on the horizon, the DC holographic one point functions they give need no renormalization. As we will see on the next sections, however, anomalous theories allow for the definition of multiple current operators depending on the physical properties they are asked to satisfy. From this point of view, the conservation law will provide a ``preferred choice'' of current from an RG perspective.

\section{Anomalous membrane currents and the role of extrinsic curvature}\label{sectionmembranecurrents}

We now turn our attention to the body of these notes, in which we analyze how the membrane construction and the Wald charges need to be extended when dealing with anomalous theories. In particular, in this section, we use the form of the constraint equations to properly define a membrane stress tensor and currents in such a way that their anomalous Ward identities coincide, inasmuch as possible, with those of a standard quantum field theory with 't Hooft anomalies.

In order to do this, let us start by a brief review of the role of bulk Chern-Simons terms in the holographic description of 't Hooft anomalies. The main idea, already implemented in early applications of the AdS/CFT correspondence,  is to employ an inflow mechanism by adding to the holographic action an appropriate Chern-Simons form
\begin{equation}
S_g \to S_g + \int d^{d+1} x \sqrt{-g} I_{CS}\left[A,F,{\bf \Gamma}, \bf R \right] \, .
\end{equation}
Notice that we write the Chern-Simons action as function of the connections and the curvatures, as opposed to $I_{CS}[A,dA, {\bf\Gamma}, d {\bf\Gamma}]$, which has to be kept in mind when taking partial derivatives. 

The presence of the connections $A$ and ${\bf \Gamma}$ in the Chern-Simons term spoil the gauge and diffeomorphism invariance of the theory. This leads to the anomalous variation of the action
\begin{equation}\label{eq:anomvariation}
\bar{\delta} S_\Sigma \doteq \int_\Sigma d^d x \sqrt{-\gamma} n^\mu \left( \alpha \frac{\partial I_{CS}}{\partial A_\mu} + \Lambda \frac{\partial I_{CS}}{\partial {\bf \Gamma}_\mu}  \right) \, , 
\end{equation}
where the variation $\bar{\delta}$ is defined through
\begin{align}
 \bar{\delta} A &= d \alpha \, , \\
 \bar{\delta} {\bf \Gamma} &= d {\bf \Lambda} + [{\bf \Lambda}, {\bf\Gamma}]\, ,
\end{align}
where $\left(\Lambda\right)^\mu_\nu= \partial_\nu \xi^\mu$ is the connection's gauge parameter.

According to the descent procedure these Chern-Simons terms can be chosen to give as variations exactly the characteristic classes of the consistent anomalies of the theory (see Appendix \ref{appsubscurrentdef} for a brief review). These will appear on the right hand side of the Ward identities for the consistent membrane currents, which are just the constraint equations projected on a constant $r$ hypersurface, as we have already reviewed.

It was pointed out long ago, e.g. in \cite{Solodukhin:2005ah}, that extra care is needed when dealing with the gravitational contributions. The ADM decomposition will in general give further terms that spoil the form of the anomaly equations unless extra constraints are imposed. As far as asymptotically AdS spacetimes are concerned, however, the usual boundary conditions are sufficient to assure that no extrinsic term survives at the conformal boundary.

In this work we take an alternative point of view on the interpretation of these contributions. Away from the conformal boundary there is in general no reason to suppress the extrinsic contributions coming from the Chern-Simons action. However one should be able to define consistently a set of Ward identities for the currents on an arbitrary bulk membrane by redefining such fields.

We show how this procedure may be carried out, fixing the form of the currents and showing how it is precisely these extrinsic terms that carry the relevant information regarding anomalous transport. In what follows we will mainly focus on the specific example of four dimensional field theories but, since the equations are known to follow a general structure, we will keep a general notation, as in \cite{Azeyanagi:2014sna}.

In four dimensions the most general anomaly for systems with a $U(1)$ current comes from the Chern-Simons form
\begin{equation}
I_{CS}= \epsilon^{\mu\nu\rho\sigma\tau} A_\mu \left( \frac{\kappa}{3} F_{\nu\rho} F_{\sigma\tau} +\lambda tr\left(\bf R_{\nu\rho} \bf R_{\sigma \tau} \right)\right) \, , \label{eq:CSaction}
\end{equation}
where we have fixed the prefactor in order to match \eqref{anomalyinintroduction}. While the first term describes a pure $U(1)^3$ anomaly, the second one gives a mixed anomaly between the current and the stress tensor. As it is well known, one can move this anomaly from one sector to the other by the addition of a total derivative, to have
\begin{equation}
I'_{CS}= \frac{\kappa}{3}\epsilon^{\mu\nu\rho\sigma\tau} A_\mu  F_{\nu\rho} F_{\sigma\tau} - 2\alpha \epsilon^{\mu\nu\rho\sigma\tau} F_{\mu\nu}  tr \left( {\bf \Gamma}_\rho \partial_\sigma {\bf \Gamma}_\tau +\frac{2}{3} {\bf \Gamma}_\rho {\bf \Gamma}_\sigma {\bf \Gamma}_\tau \right) \, 
\end{equation}
or in terms of the curvatures
\begin{equation}
I'_{CS}= \frac{\kappa}{3}\epsilon^{\mu\nu\rho\sigma\tau} A_\mu  F_{\nu\rho} F_{\sigma\tau}  - \lambda \epsilon^{\mu\nu\rho\sigma\tau} F_{\mu\nu} tr \left({\bf \Gamma}_\rho {\bf R}_{\sigma \tau}  - \frac{2}{3}{\bf \Gamma}_\rho {\bf \Gamma}_\sigma {\bf \Gamma}_\tau    \right) \, .  \label{eq:CSactionII}
\end{equation}
This has no effect on the bulk dynamics (and with it the construction of the conserved charges), and the only way in which it modifies the theory is through a redefinition of the consistent current operators. While the physical phenomena will be the same in both descriptions, their interpretation is slightly different in the two cases. For our purposes, for example, the Wald construction will be much more transparent when the anomaly is in the diffemorphism sector.\footnote{In particular this proves extremely useful when dealing with constant magnetic fields, as the linear dependence on $A_\mu$ on $I_{CS}$ makes it tricky to apply the Stokes theorem when needed. A further advantage comes at a formal level, as if $I'_{CS}$ is chosen one can move the mixed anomaly between the two sectors by the usual Bardeen counterterm, which only involves the intrinsic geometry on $\Sigma$.} Thus we will in general consider \eqref{eq:CSactionII} as the Chern-Simons action. We will however comment on the changes to the construction if \eqref{eq:CSaction} is considered instead, as it is often the case in the literature. In what follows we introduce the full bulk theory, commenting on the universality of the construction and the due caveats.

\subsection{Gravitational theory}

We will be interested in $d+1$ dimensional bulk theories of the form
\begin{equation}
S = \frac{1}{8\pi G} \int d^{d+1}x \sqrt{-g} \left(R -2\Lambda\right) + \int d^{d+1} x \sqrt{-g} L_{mat} + \int d^{d+1} x \sqrt{-g} I_{CS}[A,{\bf \Gamma}] \, , \label{IIIlagrangian}
\end{equation}
where $L_{mat}$ denotes the matter Lagrangian and $I_{CS}$ is the bulk Chern-Simons action defined before. Throughout our considerations we will take the matter Lagrangian to be a function only of the $U(1)$ curvature $F_{\mu\nu}$ and possibly uncharged scalar fields. The simplest choice is of course that of Maxwell's theory, but various generalizations have been discussed in the literature. The presence of charged matter would spoil the simple RG structure of the long wavelength limit of the theory, as charged ``hair'' would give corrections to the DC properties which account for the charge distribution throughout the bulk spacetime. As far as anomalous response goes, however, our minimal assumptions will assure the universality of the results.
 
At this point it is simple to derive the form of the bulk equations of motion, as it was done for example by \citep{Azeyanagi:2013xea}
\begin{align}
2\nabla_\mu \frac{\partial L_{mat}}{\partial F_{\mu\nu}} &= \Sigma^\nu_A \, , \label{eq:eomanomalyU1} \\ 
G^{\mu\nu} - g^{\mu\nu} \Lambda &=T_{mat}^{\mu\nu} + \nabla_\rho \Sigma^{(\mu\nu) \rho} \, ,\label{eq:eomanomalygrav}
\end{align}
where $T^{\mu\nu}_{mat}$ is the matter stress tensor and $\Sigma^{\mu\nu \rho}$ and $\Sigma_A^\mu$ are the spin and Hall currents. It will be useful to define them through bulk variations. In this case let  ${E^{\mu\nu}}_\rho$ be the Euler-Lagrange equations for the connection $\Gamma_{\mu\nu}^\rho$ as an independent field
\begin{equation}
{E^{\mu\nu}}_\rho= \left(\frac{\partial}{\partial \Gamma_{\mu\nu}^\rho} - 2 \nabla_\sigma  \frac{\partial}{\partial {{R_{\sigma \mu}}^\rho}_\nu } \right) I_{CS} \, .
\end{equation} 
Then,
\begin{align}
\Sigma^{\mu\nu\rho} &= \frac{1}{2} \left( E^{\mu\nu\rho} + E^{\mu\rho\nu} - E^{\rho\mu\nu} \right) \, , \\
\Sigma_A^\mu &=   \frac{\partial I_{CS}}{\partial A_\mu} -2 \nabla_\nu \frac{\partial I_{CS}}{\partial F_{\nu\mu}} \, .
\end{align}
In the five dimensional theory this reduces to
\begin{align}
\Sigma^{\mu\nu\rho} &= 2 \lambda \epsilon^{\mu\alpha\beta \gamma \delta} F_{\alpha \beta} {R^{\nu\rho}}_{\gamma \delta} \, , \\
\Sigma_A^\mu &= \kappa \epsilon^{\mu\alpha\beta\gamma\delta} F_{\alpha\beta} F_{\gamma \delta} \, .
\end{align}
In general both the spin and Hall currents are covariant expressions even if they come from a non-covariant action. This is what motivated a series of works on the definition of the entropy of such theories and the extension of the Wald construction to theories with a noncovariant Lagrangian \cite{Tachikawa:2006sz, Azeyanagi:2014sna, Bonora:2011gz}.

Their construction includes a large part of the technical tools we will be using in the next section and was key to us in order to extend our results past perturbation theory. We will review it in the next section, deriving the explicit formula for the Komar charge in the cases of interest.

Finally, let us point out that, if gravitational Chern-Simons terms are present, the theory will not in general have a well defined variational problem due to the presence of higher derivatives. This is of course an extremely problematic point as far as the space of solutions is concerned. It is usually justified in the context of holography by noticing that the gravitational anomaly coefficient $\lambda$ is subleading in the large $N$ expansion, and that the variational problem may get corrected by other equally subleading terms. Another way of circumventing this problem, at least as far as we are concerned, is to impose the spacetime to be asymptotically AdS from the beginning, but work remains to be done regarding the stability of such solutions.

Related issues have been studied in AdS/CFT, notably by \cite{Skenderis:2009nt} in the context of topologically massive gravity (TMG) while \cite{Liu:2016hqb} showed the presence of finite momentum instabilities for charged black-holes in five dimensions. We will comment briefly on how the first of the two studies can be linked to the modifications of the membrane currents in appendix \ref{appthermalhall}.

\subsection{Analysis of the constraint equations}

We look at the variation of the on-shell action to derive the explicit form of the constraint equations. In these theories in general no counterterm \`a la Gibbons-Hawking exists in order to make the variational problem well defined\footnote{This can have, apart from the usual Ostrogradski type instabilities, serious effects on the unitarity of the dual theories, which are still not completely clarified. See e.g. \cite{Skenderis:2009nt} for the study of the $AdS_3$ case.} and therefore the on-shell variation takes the general form
\begin{equation}
\delta  S_\Sigma = \int_\Sigma d^d x \sqrt{-\gamma}\left( \frac{1}{2} t^{ab} \delta \gamma_{ab} + \frac{1}{2} u^{ab} \delta K_{ab} + {l^{ab}}_c \delta \hat{\Gamma}_{ab}^c + \mathcal{J}^a \delta A_a \right) \, , \label{eq:onshellvariationanomalies}
\end{equation} 
where the precise form of the various quantities depends on the choice of Chern-Simons action. We separate the variation with respect to the connection from the one with respect to the induced metric as ${l^{ab}}_c$ is in general not covariant if anomalies are present. On the other hand, $t^{ab}$ only contains covariant and gauge invariant quantities.

\paragraph{Gauge constraint.}
We start by analyzing the gauge constraint. As of now the \emph{consistent} current $\mathcal{J}^a$ is simply given by
\begin{equation}
\mathcal{J}^a \doteq \frac{1}{\sqrt{-\gamma}}\frac{\delta S_g}{\delta A_a} = 2 n_\mu\left(\frac{\partial L_{\rm{mat}}}{\partial F_{\mu\nu}} +\frac{\partial I_{CS}}{\partial F_{\mu\nu}}\right) P_\nu^a \, . \label{eq:consistentcurrentanom}
\end{equation}
Integrating by parts the gauge variation of \eqref{eq:onshellvariationanomalies} and using \eqref{eq:anomvariation} we get the anomalous constraint
\begin{equation}
D_a \mathcal{J}^a = - n_\mu\frac{\partial I_{CS}}{\partial A_\mu} \, . \label{eq:currentconstraint}
\end{equation}
This constraint has a slightly different interpretation depending on whether a possible gravitational anomaly lies in the diffeomorphism or gauge sector. In the first case no gravitational characteristic class appears on the right hand side of \eqref{eq:currentconstraint} but the definition of the consistent membrane current picks up a further contribution coming from the Chern Simons term
\begin{equation}
\mathcal{J}^a= 2 n_\mu\frac{\partial L_{\rm{mat}}}{\partial F_{\mu\nu}} P_\nu^a + J_{BZ}^a + J_{CSK}^a  \, ,
\end{equation}
where the extrinsic Chern-Simons current $J_{CSK}^a$ comes from the ADM decomposition of \eqref{eq:consistentcurrentanom} and $J_{BZ}^a$ is the usual Bardeen-Zumino polynomial needed to define covariant currents. It is obtained by the purely intrinsic part of the ADM decomposition of \eqref{eq:consistentcurrentanom}. For the five dimensional theory with mixed anomaly, setting $J_{BZ}^a= J_A^a + J_{\bf{\hat{\Gamma}}}^a$, they are given by
\begin{align}
J_{CSK}^a &=  -8 \lambda \epsilon^{abcd} K_b^f D_c K_d^f \, , \\
J^a_{\bf\hat{\Gamma}}&= 4 \lambda \epsilon^{abcd} tr \left({\bf \hat{\Gamma}_b}\partial_c {\bf \hat{\Gamma}_d} +\frac{2}{3} {\bf \hat{\Gamma}_b}{\bf \hat{\Gamma}_c}{\bf \hat{\Gamma}_d}  \right) \, , \\
J_A^a &= \frac{4}{3}\kappa \epsilon^{abcd}A_b F_{cd} \, .
\end{align}
Notice that the extrinsic current is only first order in the spacetime derivatives on $\Sigma$ and thus is a good candidate to encode the anomalous horizon fluctuations related to the gravitational anomaly. A simple asymptotic analysis shows such a current to vanish identically at the conformal boundary, as shown by \cite{Azeyanagi:2014sna}, so that any relevant effect is ``dynamically'' generated as the horizon is approached. Furthermore, it is very important to appreciate that the current $J_{CSK}^a$ is a perfectly covariant and gauge invariant object from the point of view of the membrane $\Sigma$. Thus, such current cannot contribute in any way to the presence of an anomaly but it has physical consequences on the observables.

The mixed anomaly can then be moved to the gauge sector by the Bardeen counterterm
\begin{equation}
B_\Sigma= -\int_\Sigma d^d x \sqrt{-\gamma} A_a J_{{\bf \hat{\Gamma}}}^a \, , \label{Bardeencounterterm}
\end{equation}
in which case the consistent current loses its $ J_{{\bf \hat{\Gamma}}}^a$  contribution, which gives rise to the mixed anomaly according to
\begin{equation}
D_a J_{{\bf \Gamma}}^a = \lambda \epsilon^{abcd} tr \left( \bf{\hat{R}}_{ab} \bf{\hat{R}}_{cd} \right) \, .
\end{equation}
Notice how extrinsic part remains as a defining feature of the membrane current.

In studies of this theory it is more common, however, to start with a bulk Chern-Simons term which is diffeomorphism invariant, e.g. in \cite{Landsteiner:2011iq}. In this case the right hand side of the constraint equation reads
\begin{equation}
n_\mu\frac{\partial I_{CS}}{\partial A_\mu}|_{grav} = - \lambda \epsilon^{abcd} tr \left( \bf{\hat{R}}_{ab} \bf{\hat{R}}_{cd} \right) - D_a J^a_{CSK} \, ,
\end{equation}
and one has to add a counterterm explicitly dependent on the extrinsic data 
\begin{equation}
S_{CSK}= \int_{\Sigma} d^d x \sqrt{-\gamma} A_a J_{CSK}^a \, , \label{IIIcounterterm}
\end{equation} 
in order to recover the correct Ward identity and obtain the correct definition of the membrane currents.

Without the inclusion of said counterterm the horizon physics will be the same, although one would explain the anomalous fluctuations as a response to the ``thermal'' anomaly
\begin{equation}
\mathcal{A}_{thermal} = 64 \pi^2 T^2 E_g \cdot B_g \, ,
\end{equation}
where $E_g$ and $B_g$ are the gravitoelectric and gravitomagnetic fields on the horizon. See for example \citep{Chapman:2012my}. We prefer not to think about it this way, since this is not a true anomaly, as it comes from the divergence of a physical current and it is state dependent.

At this point we can go back to the choice \eqref{eq:CSactionII} for the Chern-Simons term. The standard QFT Ward identities are recovered on an arbitrary slice $\Sigma$ and one can mimic the construction of appendix \ref{appsubscurrentdef} to define all the different membrane currents. Explicitly it is possible to define consistent $\mathcal{J}^a$, covariant $J^a$ and conserved $J^a_{cons}$ currents by the defining equations
\begin{align}
\mathcal{J}^a &\doteq \frac{1}{\sqrt{-\gamma}} \frac{\delta S_g}{\delta A_a} \, , \\
\delta_{\alpha,\Lambda} J^a &= 0 \, , \\
D_a J^a_{cons} &= 0 \, .
\end{align}
The explicit expressions, letting a Bardeen counterterm $-c B_\Sigma$ be used to define the consistent current, read
\begin{align}
\mathcal{J}^a &=2 n_\mu\frac{\partial L_{\rm{mat}}}{\partial F_{\mu\nu}}  + J^a_{CSK} + J_A^a + (1-c) J^a_{\bf{\hat{\Gamma}}} \, , \label{eq:conscurrentanomaly} \\
J^a &= 2 n_\mu\frac{\partial L_{\rm{mat}}}{\partial F_{\mu\nu}}  P_\nu^a + J_{CSK}^a  \, , \\
J^a_{cons} &=J^a + J^a_{\bf{\hat{\Gamma}}} + \frac{3}{2} J_A^a \, . 
\end{align}
The factor $3/2$ depends on the fact that we are considering 4-dimension theory. It general it can be extracted by writing the $U(1)^3$ anomaly as a total divergence.

\paragraph{Diffeomorphism constraint}

We now move to the diffeomorphism constraints, which will allow us to define a consistent membrane stress tensor. We expect such construction to be potentially problematic, as the presence of higher derivative terms will spoil the form of the diffeomorphism constraint away from the conformal boundary, where most of the troublesome terms vanish  due to the AdS asymptotics.

After a counterterm proportional to \eqref{Bardeencounterterm} has been added to the action we can rearrange the on-shell variation for a diffeomorphism by using standard formulas for the Lie derivatives
\begin{align}
\mathcal{L}_\xi \gamma_{ab}&= D_a \xi_b + D_b \xi_a \, , \\ 
\mathcal{L}_\xi A_a &= \partial_a \left( A_b \xi^b \right) + F_{ab} \xi^b  \, , \\ 
\mathcal{L}_\xi K_{ab}&= \xi^c D_c K_{ab} + D_a \xi^c K_{cb} + D_b \xi^c K_{ac} \, .
\end{align}
This allows us to extract the term proportional to the Killing field $\xi^a$ from the action variation to get the diffeomorphism constraints
\begin{equation}
\delta S_\Sigma= \int_\Sigma d^d x \sqrt{-\gamma} \left( D_a \Theta^{ab} - F^{bc} \mathcal{J}_c + A^b D_c \mathcal{J}^c+ \Delta^b \right) \xi_b \, . \label{eq:actiondiffeovariation}
\end{equation}
where we have defined for convenience
\begin{align}
\Theta^{ab} &=t^{ab} + u^{ac} K_c^b + D_d \left( l^{d(ba)} + l^{(adb)} - l^{(ab)d} \right) \, , \\
\Delta^b &= - \frac{1}{2} u^{ac} D^b K_{ac} \, , 
\end{align}
and $\mathcal{J}^a$ is given in \eqref{eq:conscurrentanomaly}. Furthermore, if we divide the stress tensor into the Einstein-Hilbert and anomalous parts, $t^{ab}=t_{EH}^{ab} + t_\lambda^{ab}$, we get the following explicit expressions for $t_\lambda^{ab}$ and $u^{ab}$
\begin{align}
t_\lambda^{ab}&= n_\mu \left(\Sigma^{ab\mu} + \Sigma^{ba\mu} \right)   \, ,\label{eq:tlambda} \\
u^{ab}&= 2 n^\mu n^\nu \left( \frac{\partial I_{CS}}{\partial {R_{\mu a \nu}}^b} - \frac{\partial I_{CS}}{\partial {{R_{\mu a}}^b}_\nu} \right) \, , \\
{l^{ab}}_c &= 2 n^\nu \frac{\partial I_{CS}}{\partial {{R_{\nu a }}^c}_b } \, , \label{eq:labc}
\end{align}
which follow from the ADM decomposition of the radial component of the presymplectic current \eqref{presymplecticcurrent}. The explicit ADM form, for the choice of the Chern-Simons action \eqref{eq:CSactionII}, is
\begin{align}
 t_{EH}^{ab} &=  - \frac{1}{8\pi G} ( K^{ab} - K \gamma^{ab} ) \, , \\
 t_\lambda^{ab} &=- 8  \lambda\epsilon^{mnp(a} \left( 2 D_n K_p^{b)} F_{rm} + \gamma^{b)l} \dot{K}_{ln} F_{pm} - F_{pm} K_l^{b)} K_n^l \right) \, , \\
 u^{ab} &= 8 \lambda \epsilon^{mnp(a} F_{mn} K_p^{b)} \, , \\
 {l^{ab}}_c &= 2 \lambda \epsilon^{amnp} F_{mn} \Gamma_{pc}^b \, .
\end{align}
If a Bardeen counterterm proportional to \eqref{Bardeencounterterm} is added to the on-shell action, this only affects, apart from the consistent current, $\Theta^{ab}$ in the following way
\begin{equation}
\Theta^{ab} \to \Theta^{ab}  -c D_d \left( l^{d(ba)} + l^{(adb)} - l^{(ab)d} \right)  -4c \lambda  \epsilon^{mnp(a} D_e\left( A_m {R_{np}}^{b) e} \right) \, .
\end{equation}

Finally, the constraint equations are derived by expressing $\delta_\xi S_\Sigma$ in \eqref{eq:actiondiffeovariation} as the consistent diffeomorphism anomaly $\mathcal{A}^b$. The constraints will take the general form
\begin{equation}
C^b= D_a \Theta^{ab} - F^{bc}\mathcal{J}_c +A^b D_c \mathcal{J}^c + \Delta^b - (1-c)\mathcal{A}^b  \doteq 0 \, . \label{eq:diffeoconstranom}
\end{equation}
For our case of interest the anomaly reads
\begin{equation}
\mathcal{A}^b =  2\lambda \gamma^{bc} \frac{1}{\sqrt{-\gamma}}\partial_a \left(\sqrt{-\gamma} \epsilon^{defg}F_{de}\partial_f\Gamma_{gc}^a \right) \, .
\end{equation}
One can re-express the Ward identity in a counterterm independent form by introducing a covariant stress tensor $T^{ab}$, as in appendix \ref{appsubscurrentdef}, analogous to the previous covariant current $J^a$. This gives
\begin{equation}
T^{ab}=\Theta^{ab} + 4 c \lambda  \epsilon^{mnp(a} D_e\left( A_m {R_{np}}^{b) e} \right) + (1-c)  D_d \left( l^{d(ba)} + l^{(adb)} - l^{(ab)d} \right) \, .
\end{equation}
and the Ward identity
\begin{equation}
D_a T^{ab} - F^{bc} J_c + \Delta^b + 2\lambda \epsilon^{cdef} D_a \left( F_{cd} {R^{ab}}_{ef} \right) =0 \, .
\end{equation}
From this one can finally define a conserved stress tensor $T_{cons}^{ab}$ by including the last total derivative
\begin{equation}
T_{cons}^{ab}= T_{ab} + 2 \lambda\epsilon^{cdef}  F_{cd} {R^{ab}}_{ef} \, ,
\end{equation}
which presents a Lorentz anomaly.

For our purposes, however, it will prove more convenient to work with the heat current, which is the object that naturally appears in the conserved flux associated to diffeomorphisms. This is given by
\begin{equation}
 Q^a = {\Theta^a}_b \xi^b + A_b \xi^b {\cal J}^a \, ,
\end{equation}
and its conservation is only spoiled by the diffeomorphism anomaly
\begin{equation}
D_a Q^a = (1-c) \mathcal{A}_b \xi^b \, .
\end{equation}
Notice that the heat current we have defined must contain the extrinsic term $u^{ac} K_{cb}\xi^b$ in order to get the right conservation equation up to the anomaly. It is this term the one that allows us to recover the $\Delta^b$ term in the constraint equation. This follows from the fact that $\xi^a$ is Killing, as
\begin{equation}
\begin{aligned}
D_a \Theta^{ab} \xi_b + \Delta^b \xi_b &= D_a \left( \Theta^{ab} \xi_b\right) - \frac{1}{2} \left( u^{ac} {K^b}_c D_a \xi_b - u^{bc} {K^a}_c  D_a \xi_b   + u^{ac} \xi^b D_b K_{ab}   \right) \\
&=  D_a \left( \Theta^{ab} \xi_b\right) - \frac{1}{2} u^{ac} \mathcal{L}_\xi K_{ac} \, .
\end{aligned}
\end{equation} 
This should be viewed as a consistency check for the proposed form of the membrane stress tensor.

We can define a {\it conserved} heat current $Q_{cons}^a$ which is divergence-free. The explicit construction is found in appendix \ref{appsubscurrentdef} and the result reads
\begin{equation}
Q^a_{cons} = {\Theta^a}_b \xi^b + A_c \xi^c \mathcal{J}^a - (1-c) \frac{1}{\sqrt{-\gamma}} \partial_b \left( \sqrt{-\gamma} {l^{ba}}_c \xi^c \right) + 2 ( 1 - c) {l^{(ab)}}_c \Lambda_b^c \, .
\end{equation}
It is useful to keep in mind that we will mainly work setting $c=0$ in the formulas above.

Let us now comment on the form of the constraint equations \eqref{eq:diffeoconstranom}. First, notice that our candidate $\Theta^{ab}$ for the membrane stress tensor does not coincide with the Brown-York prescription
\begin{equation}
\Theta^{ab} \neq \frac{2}{\sqrt{-\gamma}} \frac{\delta S_g}{\delta \gamma_{ab}} \, ,
\end{equation}
because of the additional term coming from the extrinsic curvature contributions. One may wonder if this is an artifact of the way in which we have organized the various fields in the constraint equation, since one could in principle include the modification into a redefinition of $\Delta^b$. An argument against this is that it is the quantity that enters in the heat current.

Another important consistency check for our proposal is that indeed it gives the right near boundary limit for the dual stress tensor $\langle {T^a}_b \rangle$. This is easily computed from \eqref{eq:onshellvariationanomalies} as $\delta {K^a}_b= \delta {\gamma^a}_b + \ldots $ where the dots denote subleading terms. No other contributions arise as long as strictly AdS boundary conditions are imposed
\begin{equation}
\langle {T^a}_b \rangle = \lim_{r \to \infty} \sqrt{-\gamma} \left( {t^{a}}_b + {u^{a}}_b \right) = \lim_{r \to \infty} \sqrt{-\gamma} \left( t^{ab} + u^{ac} K_{cb} \right)  = \lim_{r \to \infty} \sqrt{-\gamma}{\Theta^a}_b \, .
\end{equation}
Counterterms stemming from holographic renormalization will be of no concern in our construction. 

By moving the anomalies entirely to the gauge sector one can check by using the asymptotic expansion of the metric that the result for the bare stress tensor from \citep{Megias:2013joa} is recovered
\begin{equation}
\begin{aligned}
\langle {\mathcal{T}^a}_b\rangle &=\lim_{r \to \infty} \bigg( \frac{1}{8\pi G} \sqrt{-\gamma} \left( {K^a}_b - {\delta^a}_b K \right) \\
& + 4 \lambda \sqrt{-\gamma} \epsilon^{mnp(a} \left( -2 D_e \left( {R_{npb)}}^e A_m \right) + \frac{1}{2}F_{mn} R_{pb)} \right)  \bigg)\, .
 \end{aligned}
\end{equation}

Our prescription has been applied in systems where momentum relaxation is introduced with success. In this case, the higher derivative corrections stemming from $u^{ab}$ are crucial for the restoration of the symmetry of the mixed two point functions in the presence of a magnetic field \cite{Copetti:2017ywz}.

However it should not be forgotten that the constraint equations \eqref{eq:diffeoconstranom} still differ in general from the usual Ward identities away from the conformal boundary, where the $\Delta_b$ term vanishes guaranteeing consistency with the QFT Ward identities. Such difference takes a suggestive form if we think of $u^{ab}$ as the expectation value of an independent operator, associated with the mode excited by the extrinsic curvature. This is similar to the reasoning in \citep{Skenderis:2009nt}, where, in the three dimensional theory, such operator was indeed shown to survive at the conformal boundary for general solutions and lead to logarithmic correlators with the canonical stress tensor.

From our point of view this is reflected in the constraint equation, where 
\begin{equation}
\Delta_b \sim \mathcal{O}^{ab} D_b \Phi_{ab}
\end{equation}
appears in the form {\it operator} times $\partial(source)$ and suggests that this second operator, even if turned off at the conformal boundary, dynamically gets an expectation value as we slide through the bulk. We will try to expand this similarity in appendix \ref{appthermalhall}.

In the following section we will match the membrane currents defined from the constraint equations to the conserved fluxes related to bulk diffeomorphisms and gauge transformations. This will allow to make explicit our intuition about the anomalous membrane currents describing anomalous hydrodynamic modes by looking at the conserved fluxes in the near horizon limit.

\section{Conserved fluxes and transport coefficients}

We now proceed to construct the conserved fluxes associated with diffeomorphisms and gauge transformations for the Chern-Simons theories considered. This is largely taken from \cite{Azeyanagi:2014sna, Bonora:2011gz} where the subtleties of the construction are also discussed at length. It will be important to keep in mind that all the constructions can be derived by explicitly using the equations of motion \eqref{eq:eomanomalyU1} and \eqref{eq:eomanomalygrav} and assuming the background solution to posses a Killing vector field. This is important because of two reasons: first, it makes the construction independent of the choice of counterterms, which, for example, can be used to shift mixed anomalies between different sectors. Second, for the same reason, it will assure the charges to be independent of any holographic renormalization procedure to remove divergences.

\subsection{Construction of the anomalous Komar charges}
In the presence of bulk Chern-Simons terms the Wald construction is known to be plagued by ambiguities \citep{Bonora:2011gz}. These are due to the lack of covariance for the bulk action, which introduces various subtleties in the extraction of the Komar form $k^{\mu\nu}$ from the Noether charge. In particular, the Chern-Simons terms give further contributions to the on-shell vanishing Noether current $J_{\xi,\alpha}$ which are proportional to the gauge parameters $\alpha$ and $\Lambda$.  One can however still derive the conservation equation for the appropriate fluxes once a particular gauge is chosen. The charges thus derived are not covariant, but a covariant prescription for the differential Noether charge was given in \citep{Azeyanagi:2014sna}, which allows one to unambiguously define the Wald entropy at a bifurcation surface. We will follow instead the ideas of \citep{Bonora:2011gz} and the remarks in (5.3) of \citep{Azeyanagi:2014sna} which are closer in notation to the construction of section 2. The price to pay will be a noncovariant expression for $k^{\mu\nu}$, which from our perspective is a feature rather than a bug. In fact, it allows us to link the flux conservation to the RG properties of \emph{conserved} currents.

As we have already introduced the basic construction in section 2 we will be brief and point out only the important differences. As before, the variation of the Lagrangian, supplemented with the appropriate Chern-Simons term, may be written as
\begin{equation}
\delta \left( L+I_{CS} \right) = E \delta \Phi + d \theta \, .
\end{equation}
However, this time, when the variation is taken with respect to a pair of diffeomorphisms and gauge transformations $(\xi, \alpha)$ the Lagrangian does not change simply as a Lie derivative due to the lack of covariance of the Chern-Simons terms. Indeed a further piece $\Xi_{\xi,\alpha}$ arises because of the inflow of the consistent anomalies
\begin{align}
\delta_{\xi,\alpha} L &= d i_\xi L \, , \\
\delta_{\xi,\alpha} I_{CS} &= d i_\xi I_{CS} + d \Xi_{\xi,\alpha} \, ,
\end{align}
where 
\begin{equation}
\Xi_{\xi,\alpha} = \alpha \frac{\partial I_{CS}}{\partial A} + \Lambda \frac{\partial I_{CS}}{\partial {\bf \Gamma}}
\end{equation}
contains the consistent anomaly and, as before, $\left( \Lambda \right)^a_b = \partial_b \xi^a$. As the new contribution appears as a total derivative it is still possible to define an on-shell closed Noether current. It now reads
\begin{equation}
J_{\xi,\alpha}= \theta_{\xi,\alpha} - i_\xi \left( L + I_{CS} \right) - \Xi_{\xi,\alpha} \, , 
\end{equation}
and it is, as before, closed on-shell 
\begin{equation}
d J_{\xi,\alpha} \doteq 0 \, .
\end{equation}
In the same spirit as that of section 2 we now define by the addition of a total derivative an improved current $\hat{J}_{\xi,\alpha}$ which vanishes on-shell. We then would like to use this fact to prove that this implies the existence of conserved fluxes once the gauge transformation is chosen to preserve the solution. While the reasoning goes through unchanged for most of the process, the anomalous piece $\Xi_{\xi,\alpha}$ does, as anticipated, present some subtleties because of its proportionality to the gauge parameters. We circumvent the problem by introducing a $d-1$ form $y$ defined through the condition
\begin{equation}
\Xi_{\xi,\alpha} - dy = \Xi'_{d\alpha, d\Lambda} \, ,
\end{equation}
where $\Xi'$ only depends on the gauge parameters through their exterior derivative.
The existence of a solution to this equation is assured by the closedness of the anomaly polynomials. Summing and subtracting $y$ to the on-shell vanishing charge, we get
\begin{equation}
\hat{J}_{\xi,\alpha} = \theta_{\xi, \alpha} - i_\xi(L + I_{CS}) - \Xi'_{d\alpha, d\Lambda} + d k'_{\xi ,\alpha} \, ,
\end{equation}
where
\begin{equation}
k'_{\xi, \alpha}= k_{\xi,\alpha} - y
\end{equation}
is the new candidate for the Komar form. Notice that now the remaining part of the on-shell vanishing current is only proportional to $d\alpha, \ d\Lambda$. In the case in which we choose $d\alpha=d\Lambda=0$, such term vanishes. This allows us, if we take $\alpha=const$ and $\xi^a$ to be Killing and constant, as in section 2, to obtain the desired closure relation back
\begin{equation}
 dk_{\xi,\alpha}' =  0 \, .
\end{equation}
More details on the generality and coordinate dependence of this construction can be found in \cite{Bonora:2011gz}. Let us work out in more detail the examples that will be relevant to connect the conserved fluxes to the membrane currents.

\paragraph{Anomalous gauge charge}
In the anomalous case the form $\theta$ can be read from appendix \ref{AIIpresCS}:
\begin{equation}
\theta_\alpha^\mu = 2\left( \frac{\partial L}{\partial F_{\mu\nu}}+ \frac{\partial I_{CS}}{\partial F_{\mu\nu}} \right)\partial_\nu \alpha \, .
\end{equation}
On the other hand, the anomalous term is given by
\begin{equation}
\Xi_\alpha^\mu = \alpha \frac{\partial I_{CS}}{\partial A_\mu}\, ,
\end{equation}
while the Noether charge is the sum of the two contributions
\begin{equation}
J_\alpha = \theta_\alpha - \Xi_\alpha \, .
\end{equation}
Now, in order to make this current vanish on-shell, we need to add a surface term 
\begin{equation}
k^{\mu\nu} = - 2 \alpha \left(\frac{\partial L}{\partial F_{\mu\nu}}+ \frac{\partial I_{CS}}{\partial F_{\mu\nu}} \right) \, ,
\end{equation}
through its derivative $\nabla_\nu k^{\mu\nu}$. This gives the following Noether current
\begin{equation}
\hat{J}^\mu_\alpha= \alpha\left( -2 \nabla_\nu \frac{\partial L}{\partial F_{\mu\nu}} -2 \nabla_\nu \frac{\partial I_{CS}}{\partial F_{\mu\nu}} - \frac{\partial I_{CS}}{\partial A_\mu}\right) = - \alpha \left( 2  \nabla_\nu \frac{\partial L}{\partial F_{\mu\nu}} + \Sigma_A^\mu \right) \, ,
\end{equation}
which vanishes from \eqref{eq:eomanomalyU1}. Now it only remains to manipulate the consistent anomaly in order to solve the equation
\begin{equation}
\alpha \frac{\partial I_{CS}}{\partial A_\mu} - \nabla_\nu y^{\mu\nu} = \Xi'^{\mu\nu} \nabla_\nu \alpha \, .
\end{equation}
There exists a two-form $\tilde{J}^{\mu\nu}$ that satisfies
\begin{equation}
\frac{\partial I_{CS}}{\partial A_\mu}= - \nabla_\nu \tilde{J}^{\mu\nu} \, . \label{eq:tildeJmunu}
\end{equation}
With this, we can fix the form $y$ in the gauge case to be
\begin{equation}
y^{\mu\nu}= \alpha \tilde{J}^{\mu\nu} \, . 
\end{equation}
Notice that this is just the construction required in order to define a \emph{conserved} current from a consistent one. For example, in 4 dimensions one gets
\begin{equation}
\frac{1}{2} J_A^a = n_\mu \tilde{J}^{\mu\nu} P_\nu^a \, .
\end{equation}
We can use this to identify the current appearing in the Komar two form for the anomalous gauge transformations
\begin{equation}
k'^{\mu\nu} =  -2 \left(\frac{\partial L}{\partial F_{\mu\nu}}+ \frac{\partial I_{CS}}{\partial F_{\mu\nu}} \right) -\tilde{J}^{\mu\nu} \, .
\end{equation}
It gives 
\begin{equation}
n_\mu k'^{\mu\nu} P_\nu^a = J^a_{cons} \, .
\end{equation}
If the fields decay fast enough at the boundary of $\Sigma$, the RG conservation for the flux of the gauge charge on $\Sigma$ can be reinterpreted as the conservation of the flux of the \emph{conserved} membrane current $J_{cons}^a$
\begin{equation}
\partial_r \int_\Sigma \sqrt{-\gamma} J^a_{cons}=0 \, .
\end{equation}
However, for particular cases like the $U(1)^3$ anomaly, the continuity equation of the Komar form doesn't reduce simply to the radial conservation of the flux of the conserved membrane currents. It will also have a surface contribution. In fact, it was shown by \cite{Grozdanov:2016ala} that for the $U(1)^3$ anomaly it occurs when a constant magnetic field is present. 

Now the lack of covariance of the construction is translated into the lack of covariance of $J^a_{cons}$ 
\begin{equation}
\delta_{\alpha} J^a_{cons} \neq 0 \, , \ \ \delta_{\Lambda} J^a_{cons} \neq 0 \,.
\end{equation}
We will see how this conservation can be used to recover the gravitational anomalous transport as horizon fluctuations of the extrinsic Chern-Simons currents $J_{CSK}^a$.

Let us also comment here on the results of \cite{Grozdanov:2016ala} on the universality of the anomalous $U(1)$ transport effects in holography from the point of view of the Komar construction. More precisely, the authors show that the results for the anomalous conductivities remain valid in a large number of theories in which the field strength appears in higher powers. From our Komar charge construction this is seen as a modification of the $\frac{\partial L}{\partial F}$ contribution. This vanishes at the horizon in the absence of external electric fields once infalling boundary conditions are imposed.

Furthermore, they argue that in the presence of massive vector fields in the bulk, or a St\"uckelberg field, the form of the conservation equation breaks down, and the boundary conductivities receive nontrivial bulk corrections. This is easily seen from the equations of motion, as in this case the Maxwell equation ceases to be a total derivative due to the bulk charged matter current. In its simplest realization this makes the bulk $U(1)$ field massive and the dual current to acquire an anomalous dimension and to be no longer conserved.

In our construction this happens because for St\"uckelberg fields $\theta$ the choice of a constant gauge parameter $\alpha$ does not lead to an invariant transformation as $\delta_\alpha \theta= \alpha \neq 0$. Thus there is no choice for $\alpha$ that makes the field configuration invariant.

This makes the presymplectic form nonvanishing on-shell and the Komar form fails to be closed. As a matter of fact, corrections to the axial conductivity due to bulk charged scalar fields were found in \cite{Copetti:2016ewq} and linked to the infrared screening of axial charge in QFT's with an emergent chiral symmetry.
It would be very interesting if a simple way of quantifying the amount of screening in an analytic manner could be derived for long wavelength fluctuations.

\paragraph{Anomalous diffeomorphism charge}

The construction in the case of diffeomorphisms is slightly more involved due to the higher number of terms one has to account for. Throughout the construction we will always assume mixed anomalies to lie in the diffeomorphism sector, since this is best suited for the presence of constant background magnetic fieds. The Chern-Simons action for the gauge anomalies in this case presents some subtleties due to the asymptotic divergence of the vector potential.

As before, one starts by extracting the presymplectic current $\theta_\xi$. One can decompose it into its Einstein-Maxwell part, given in \eqref{eq:thetapresymplEH}, and the Chern-Simons contribution given in appendix \ref{AIIpresCS}
\begin{equation}
\theta_\xi = \theta^{EM}_\xi  + \frac{\partial I_{CS}}{\partial F} \left( d i_\xi A + i_\xi F \right)  + \frac{\partial I_{CS}}{\partial {\bf R}} \left( di_\xi {\bf\Gamma} + i_\xi d {\bf\Gamma} + d \Lambda + [\Lambda, {\bf \Gamma}] \right) +  \Sigma^{\alpha\beta \mu} \nabla_{(\alpha} \xi_{\beta)} \star d x^\mu \, . \label{eq:thetaanomalous}
\end{equation}
After inclusion of the anomalous term the on-shell vanishing Noether current is then given by
\begin{equation}
\hat{J}_\xi = \theta_\xi - i_\xi \left( L + I_{CS} \right) - \Xi_\xi +d k_\xi \, , \label{Noethercurrent}
\end{equation}
where the total derivative of
\begin{equation}
k^{\mu\nu}= k_{EH}^{\mu\nu} -\frac{1}{2} \xi^\rho \left({{\Sigma^\mu}_\rho}^\nu + {\Sigma^{\nu\mu}}_\rho + {{\Sigma_\rho}^{\mu\nu}}\right) + \xi^\rho A_\rho \frac{\partial I_{CS}}{\partial F_{\mu\nu}} + \nabla_\rho \xi^\sigma \frac{\partial I_{CS}}{\partial {R^\sigma}_{\rho\mu\nu}} \, , \label{eq:kdiffeoanom}
\end{equation}
has been added to the action in order to satisfy the Einstein equations, as shown in \citep{Azeyanagi:2014sna}. Let us briefly see how this plays out. The Einstein-Maxwell part works exactly as in section 2, so that we would like to reproduce the Hall and spin currents in the equations of motion from the anomalous terms. The spin current is extracted from summing the last part of \eqref{eq:thetaanomalous} and the first anomalous contribution in $k^{\mu\nu}$. The other parts combine into the Maxwell equation with anomalous contributions and cancel the $i_\xi I_{CS}$ term through the usage of identities between differential forms. 

Following the previous discussion we add and subtract to \eqref{Noethercurrent} the total derivative of $y^{\mu\nu}$, so that we will have the on-shell conserved quantity
\begin{equation}
k'^{\mu\nu}= k^{\mu\nu}- y^{\mu\nu}\, .
\end{equation}
It remains to fix the $y^{\mu\nu}$ term. This can in general be done once a particular theory is fixed. General formulas for these terms where given in \cite{Bonora:2011gz}.

Let us then focus on the example we are interested in, the theory with the mixed anomaly
\footnote{In this construction we choose the anomaly to be in the diffeomorphism sector. This makes no physical difference, since the whole reasoning follows from the equations of motion which are left invariant by this change. In the case of the gauge anomaly one concludes $y^{\mu\nu}$ to vanish. However, the last term in \eqref{eq:kdiffeoanom} gives a non gauge invariant contribution
\begin{equation}
\nabla_\rho \xi^\sigma \frac{\partial I_{CS}}{\partial \bf R^\sigma_\rho} \sim \epsilon^{\mu\nu \alpha \beta \gamma} A_\gamma{ R^{\sigma \rho}}_{\alpha \beta} \nabla_\rho \xi_\sigma \, ,
\end{equation}
which is ill suited to the constant magnetic field analysis we are going to perform, as the linear coordinate dependence of the gauge connection above makes manipulations through Stokes theorem subtle and ill defined.
}
\begin{equation}
I_{CS} = \lambda F \wedge CS({\bf \Gamma}) \, . \label{IVchernsimonsterm}
\end{equation}
First we notice that
\begin{equation}
\frac{\partial I_{CS}}{\partial {\bf \Gamma}} = {\bf R} - 2 {\bf \Gamma}^2= d {\bf \Gamma} \, ,
\end{equation}
and we thus see upon a simple partial integration that this means
\begin{equation}
y^{\mu\nu} = \lambda \epsilon^{\mu\nu\rho\tau \sigma}F_{\tau \sigma} \Gamma_{\rho\beta}^\alpha \Lambda^\beta_\alpha \, . \label{IVy5d}
\end{equation}
We can now connect this quantity to the RG flow of the heat current defined from our previous membrane currents. One starts by taking a static solution in which $\xi= \partial_t$ is a Killing field and confront the expression for the charge with our previous prescription for the membrane stress tensor $\Theta^{ab}$.

We now look at the explicit form of $k^{ar}$. After performing an ADM decomposition, one identifies
\begin{align}
\left( k_{EH}^{\mu\nu} + \xi^\rho A_\rho \frac{\partial I_{CS}}{\partial F_{\mu\nu}} \right) n_\nu P_\mu^a &= - \frac{1}{2} t_{EH}^{ab} \xi_b - \frac{1}{2} A^b \xi_b {\cal J}^a \, , \\
- \frac{1}{2} \xi_b \left( \Sigma^{\mu b\nu} + \Sigma^{b \mu \nu} \right) n_\nu P_\mu^a &=  - \frac{1}{2} t_\lambda^{ab} \xi_b \, , \\
\left( - \frac{1}{2} \xi_b \Sigma^{\nu \mu b} + \nabla_\rho \xi^\sigma \frac{\partial I_{CS}}{\partial {R^\sigma}_{\rho\mu\nu}}  - y^{\mu\nu} \right) n_\nu P_\mu^a &= - \frac{1}{2} u^{ac} K_{cb} \xi^b - \frac{1}{2} \zeta^{a} \, ,
\end{align}
where $\zeta^a$ is given by
\begin{equation}
\zeta^a = 4 {l^{(ab)}}_c \Lambda_b^c + {l^{ab}}_c \hat{\Gamma}_{bd}^c \xi^d + {l^{ba}}_c \hat{\Gamma}_{bd}^c \xi^d - {l^{bc}}_d \hat{\Gamma}_{bc}^a \xi^d - 2 \frac{1}{\sqrt{-\gamma}} \partial_b \left( \sqrt{-\gamma} {l^{ba}}_c \xi^c \right) \, .
\end{equation}
The equations above look rather messy. However, we can use the intuition we have gained from the $U(1)$ case to check whether they combine into the conserved heat current
\begin{equation}
 k_\xi^{ar} = - \frac{1}{2} \left( \Theta^{ab} \xi_b + A^b \xi_b {\cal J}^a - \frac{1}{\sqrt{-\gamma}} \partial_b \left( \sqrt{-\gamma} {l^{ba}}_c \xi^c \right) + 2 {l^{(ab)}}_c \Lambda_b^c \right) \, . \label{Komarform}
\end{equation}
The computation is very tedious but conceptually straight-forward, as one needs to compute explicitly the intrinsic contributions to $\Theta^{ab}$. The computation simplifies greatly by realizing 
\begin{equation}
 {l^{ba}}_c \hat{\Gamma}_{bd}^c \xi^d = - {l^{bc}}_d \hat{\Gamma}_{bc}^a \xi^d \, .
\end{equation}
One finally finds 
\begin{equation}
\Theta^{ab} \xi_b = t_0^{ab} \xi_b + t_\lambda^{ab} \xi_b + u^{ac} K_{cb} \xi^b - \frac{1}{\sqrt{-\gamma}} \partial_b \left( \sqrt{-\gamma} {l^{ba}}_c \xi^c \right) + 2 {l^{(ab)}}_c D_b \xi^c - {l^{bc}}_d \Gamma_{bc}^a \xi^d \, ,
\end{equation}
which allows to prove \eqref{Komarform}. Then, for surface terms decaying fast enough, the conservation equation reads
\begin{equation}
0= \partial_r \int_{\Sigma} \sqrt{-\gamma}\left( \Theta^{ab} \xi_b + A^b \xi_b {\cal J}^a - \frac{1}{\sqrt{-\gamma}} \partial_b \left( \sqrt{-\gamma} {l^{ba}}_c \xi^c \right) + 2 {l^{(ab)}}_c \Lambda_b^c \right) = \partial_r \int_\Sigma \sqrt{-\gamma} Q^a_{cons} \, ,
\end{equation}
which allows us to interpret the conserved flux as an RG equation for the membrane conserved heat current. Notice, in particular, that it is $\Theta^{ab}$ and not the Brown-York tensor the one to appear in the RG equation.

\subsection{Choice of background and perturbations}

Finally we show how the formalism developed in the last section works by extracting the anomalous transport coefficients from the five dimensional Einstein-Maxwell-Chern Simons theory.

We proceed by postulating a general asymptotically AdS stationary black hole solution. Formally it should be intended as an order by order fluid-gravity expansion in spatial derivatives. The transport coefficient we will extract correspond to the first order corrections to the one point functions of the currents. However it is instructive to keep the discussion as general as possible in the near-horizon region, where only regularity and stationarity need to be imposed in order to evaluate the extrinsic membrane currents.

As we are interested in the effects of both $U(1)^3$ and mixed gauge-gravitational anomalies we study charged black hole solutions with no electric fields. Once the stationarity condition is imposed our ansatz takes the form
\begin{align}
ds^2 &= dr^2  - f(r) u_a u_b dx^a dx^b + g(r) h_{ab} dx^a dx^b \, , \\
A &= A_t (r) u_a dx^a + a_b dx^b \, .
\end{align} 
Because of the presence of the Killing vector $\xi^a= (1, 0,0,0)$, we have naturally decomposed the metric and the one-form $A$ into their projections parallel and perpendicular to $\xi^a$. In particular, we get
\begin{align}
u_a &= -\frac{1}{f(r)}\gamma_{ab} \xi^b \, ,\\
h_{ab} \xi^b &= {h_a}^b u_b = 0 \, \\
A_t &= A_a \xi^a \, , \\ 
a_b \xi^b &= 0 \, .
\end{align}
All of the functions in the ansatz above, if not indicated explicitly, depend on the radial and the spatial coordinates only.
We define the magetic field $B^a$ and the vorticity $\omega^a$ on a spacetime slice $\Sigma$ as usual through 
\begin{align}
B^a &= \sqrt{-\gamma} \epsilon^{abcd} u_b \partial_c a_d \, , \\
\omega^a &= - \frac{1}{2} \sqrt{-\gamma}\epsilon^{abcd} u_b \partial_c u_d \, . 
\end{align}
Notice that the magnetic part of the $U(1)$ field strength is not only the magnetic field. Rather it mixes with the vorticity through the time component of the gauge field 
\begin{equation}
\sqrt{-\gamma} \epsilon^{abcd} u_b \partial_c A_d= B^a - 2 A_t \omega^a \, .
\end{equation}
In order to describe an holographic system, the solution has to satisfy boundary conditions both on the horizon and on the conformal boundary. In particular, as $r \to \infty$, the asymptotically AdS structure implies
\begin{align}
f(r) &\sim g(r) \sim e^{2r} \, , \\
h_{ab} &\sim u_b \sim O(1) \, , \\
A_t &\sim a_b \sim O(1) \, . 
\end{align}
On the other hand, the horizon asymptotics are fixed by regularity and infalling conditions \citep{Iqbal:2008by}. In particular, the function $f(r)$ is vanishing at the location $r_H$ of the horizon, while its derivative is related to the Hawking temperature $T$ through
\begin{equation}
\dot{f}(r) = 4 \pi T \sqrt{f(r)} + O(f) \, .
\end{equation}
We can fix, without loss of generality, $g(r_H)=1$ and $\dot{g}(r_H)=1$, so that $h_{ab}$ plays the role of the induced metric on the horizon. For the gauge field the infalling boundary conditions and the absence of electric field $F_{ab}\xi^b$ imply 
\begin{equation}
 F^{ra}(r_H)=0 \, .
\end{equation}
In evaluating the contributions to the horizon membrane currents we will take the expansion for the extrinsic curvature near the horizon to be
\begin{equation}
K_{ab}=\frac{1}{2} \sqrt{f} \left(-4 \pi T u_a u_b + h_{ab} \right)  +O(f) \, ,
\end{equation}
and make extensive use of the norm $u_a u^a = -1/f(r)$.

Finally, in order to avoid cluttering of formulas we will fix the gauge for the vector potential in such a way that $\lim_{r \to \infty} A_t=0$ and $A_t(r_H)= -\mu$, where $\mu$ is the chemical potential of the dual CFT state. As we will deal with noncovariant charges the choice of gauge does play an important role. In particular, to work in a general gauge, one should first extract the boundary covariant currents from the Bardeen polynomials. Covariance then will assure the answer for the one point function to be expressed through the gauge invariant chemical potential
\begin{equation}
\mu = \int dr F_{ra} \xi^a = A_t(\infty) - A_t(r_H) \, .
\end{equation} 

\subsection{Membrane paradigm for anomalous currents}

We are now going to compute the transport coefficients using the ansatz presented above. We start by doing it for the case in which we have the mixed anomaly and, after that, we also include how the computations would go for the case in which we have $U(1)^3$ anomaly. In both cases, the structure followed will be presenting the conserved fluxes associated to both gauge transformations and diffeomorphisms, express it in terms of the membrane currents and then perform the computations of the 1-point functions of the field theory operators.

\subsubsection{Mixed gauge-gravitational anomaly}

Before starting we remind the reader of the precise form of the bulk Chern-Simons term we will employ
\begin{equation}
I_{CS} = \int d^5 x \sqrt{-g} 2 \lambda \epsilon^{\mu\nu\rho\sigma\tau} F_{\mu\nu} \left( \Gamma_{\rho\beta}^\alpha \partial_\sigma \Gamma_{\tau\alpha}^\beta + \frac{2}{3} \Gamma_{\rho\gamma}^\alpha \Gamma_{\sigma\alpha}^\beta \Gamma_{\tau\beta}^\gamma \right) \, . 
\end{equation}
As we add no Bardeen counterterms, this corresponds to the choice of $c=0$ in the construction of the consistent membrane currents of section 3. One can explicitly verify that, with this choice, the continuity equation $\nabla_\mu k_{\alpha,\xi}^{\mu\nu} = 0$ can be integrated on $\Sigma$ with no contributions from its boundary. Then we extract the boundary DC one point functions from the matching of the conserved fluxes $I^a$ and $H^a$.

As we have shown above, these fluxes are related to  the conserved membrane currents. We rewrite the formulas here for convenience
\begin{align}
I^a &= \int_\Sigma d^d x \sqrt{-\gamma} J^a_{cons} \, , \\
H^a &= \int_\Sigma d^d x \sqrt{-\gamma} Q^a_{cons} \, ,
\end{align}
where the explicit expressions for the integrands are
\begin{align}
J^{a}_{cons} =& \mathcal{J}^a= F^{ra} + J_{CSK}^a + J_{{\bf\hat{\Gamma}}}^a \, , \\
Q^a_{cons} =& {\Theta^a}_b \xi^b + A_c \xi^c \mathcal{J}^a - \frac{1}{\sqrt{-\gamma}} \partial_b \left( \sqrt{-\gamma} {l^{ba}}_c \xi^c \right) + 2 {l^{(ab)}}_c \Lambda_b^c \, .
\end{align}
These quantities, when evaluated on the conformal boundary, reduce to the respective CFT one point functions, as shown above.

Let us start by considering the current one point functions. Integrating the conserved flux $I^a$ in the radial direction gives
\begin{equation}
\int d^4x \sqrt{-\gamma_{(0)}} \langle J^a_{cons} \rangle = \int_H d^4x \sqrt{-\gamma} J^a_{cons} (r_H) \, .
\end{equation}
The left-hand side, as long as we are only interested in contributions which are first order in derivatives, coincides with the covariant current of the CFT, since the term $ J_{{\bf\hat{\Gamma}}}^a$ gives contributions only starting from third order. In the horizon evaluation we can discard this term for the same reason, while the non-anomalous $F^{ra}$ term vanishes identically from the infalling boundary conditions. The one point function is then expressed as the horizon integral of the extrinsic Chern-Simons current $J_{CSK}^a$
\begin{equation}
\int d^4x \sqrt{-\gamma_{(0)}} \langle J^a \rangle = \int_H d^4x \sqrt{-\gamma} J^a_{CSK} (r_H) + O(\partial^3) \, .
\end{equation} 
The value of the Chern-Simons current at the horizon is completely general and is only a consequence of the regularity of the near horizon geometry. In fact, explicitly expanding its expression
\begin{equation}
\begin{aligned}
\sqrt{-\gamma} J_{CSK}^a &= -8 \lambda \sqrt{-\gamma} \epsilon^{abcd} {K_b}^e D_c K_{de} \\
&= 32 \lambda \pi^2 T^2 \sqrt{-\gamma}\epsilon^{abcd} u_b \partial_c u_d + \lambda \sqrt{-\gamma} \epsilon^{abcd} \left( 16 \pi T u_b u^e D_c h_{de} - 2 f {h_b}^e D_c h_{de} \right) \\
&= 32 \lambda \pi^2 T^2 \sqrt{-\gamma}\epsilon^{abcd} u_b \partial_c u_d  + O(f) \, .
\end{aligned}
\end{equation}
Integrating and remembering that the vorticity zero mode is constant in the bulk, one gets the familiar answer
\begin{equation}
\int d^4x \sqrt{-\gamma_{(0)}} \langle J^a \rangle = - 64 \lambda \pi^2 T^2 \omega^a \, .
\end{equation}
We stress how, in this setting, the gravitational transport follows from the state dependent extrinsic contribution to the membrane currents given by $J_{CSK}^a$ only. It would be interesting to understand whether it is possible to link out of equilibrium fluctuations of the chiral vortical effect to horizon variations of such membrane current.

\bigskip

Let us now turn to the treatment of the diffeomorphism conserved flux. In this case, we are also interested in working only at first order in derivatives. Then, integrating the diffeomorphism flux $H^a$ between the boundary and the horizon gives, in the chosen gauge,
\begin{equation}
\int d^4 x \sqrt{-\gamma_{(0)}} \langle {T^{a}}_b \xi^b \rangle = \int_{H} d^4 x \sqrt{-\gamma} \left( {t^a}_b \xi^b + u^{ac} K_{cb} \xi^b + A_c \xi^c J_{CSK}^a  \right) + O(\partial^3) \, ,
\end{equation}
where $O(\partial^3)$ denotes all the remaining intrinsic terms which only start contributing at third order in derivatives, given that both the horizon and the boundary are flat.

We thus need to evaluate the right hand side on the horizon to extract (minus) the one point function of the energy current. The third term on the right-hand-side follows immediately from the evaluation of the gauge fluxes and gives
\begin{equation}
A_c \xi^c J^a_{CSK}= 64 \lambda \pi^2 T^2 \mu \omega^a \, .
\end{equation}
where we have used the already commented choice $A_c \xi^c(r_H)= A_t(r_H)= -\mu$.
The first term of the right-hand-side, on the other hand, can be split into the Einstein-Hilbert part \eqref{IIBYork} and the anomalous part ${ t_\lambda^a}_b$ made up of spin currents \eqref{eq:tlambda}.
The first of the two is easily evaluated to give the ideal  part of the stress tensor, which however does not contribute to the heat current due to the orthogonal projection.
The anomalous part can also be shown to vanish at the horizon:
\begin{equation}
\sqrt{-\gamma}{t_\lambda^a}_b \xi^b (r_H)= 32  \pi^2 T^2 \lambda \sqrt{-\gamma} \epsilon^{efga} \left( u_f u_b \xi^b - u_f u_b \xi^b \right) F_{ge} + O(f)=0 \, ,
\end{equation}
where we have used the asymptotic expansion of the extrinsic curvature and the vanishing of $F_{ra}$ at the horizon. Notice that these two quantities make up the Brown-York prescription for the stress tensor at the horizon, but they show no anomalous transport.

Finally it remains to evaluate the extrinsic anomalous part
\begin{equation}
\sqrt{-\gamma} u^{ab} K_{bc} \xi^c= 16 \pi^2 T^2 \lambda \epsilon^{aefg} F_{ef} u_g + O(f)= 32 \pi^2 T^2 \lambda \left( B^a + 2 \mu \omega^a \right) \, .
\end{equation}
Expressing everything together, we obtain
\begin{equation}
\int d^4 x \sqrt{-\gamma_{(0)}}\langle {T^a}_b \xi^b \rangle = 32 \pi^2 T^2 B^a + 128 \pi^2 T^2 \mu \omega^a \, .
\end{equation}
which gives the right coefficients for the anomalous gravitational transport.
Notice that now the chiral magnetic effect for the energy current, which persists without chemical potential, is completely captured by the horizon fluctuations of the extrinsic part of the modified membrane stress tensor ${\Theta^a}_b$. In this sense, this quantity is the energy counterpart of the extrinsic Chern-Simons current $J_{CSK}^a$ and we also wonder if further information about the nonequilibrium dynamics of the energy chiral magnetic effect may be extracted from the time dependent fluctuations of this quantity.

Finally, notice that, in the way in which they are presented here, these arguments can be generalized immediately to systems with momentum relaxation as in \citep{Copetti:2017ywz}. From this point of view the lack of corrections in the energy currents follows from the regular geometry of the horizon in momentum relaxation solutions.

\subsubsection{U(1)\textsuperscript{3} anomaly}
We now move to the analysis of the effect of the $U(1)^3$ anomalies on the membrane currents and its link to anomalous transport. The exact Chern-Simons action we are using for this computation is the following:
\begin{equation}
I_{CS} = \int d^{5} x \sqrt{-g} \frac{\kappa}{3} \epsilon^{\mu\nu\rho\sigma\tau} A_\mu F_{\nu\rho} F_{\sigma\tau} \, .
\end{equation}
In this case, as noticed in \citep{Grozdanov:2016ala}, further care is needed in naively integrating the continuity equation $\nabla_\mu k^{\mu\nu}=0$. In fact the linear dependence on $A_\mu$ in the expressions for $k^{\mu\nu}$ impedes the application of Stokes theorem if magnetic zero modes are present. As finite momentum modes decay faster at infinity, we may as well fix the magnetic field and the vorticity to be constant throughout the computation.

The important continuity equations now read
\begin{align}
\partial_r I^a + \int_\Sigma d^4x \left( \partial_b \sqrt{-\gamma} k^{ba}_\alpha \right) &=0 \, ,\label{eq:conservationI} \\
\partial_r H^a  + \int_\Sigma d^4x \left( \partial_b \sqrt{-\gamma} k^{ba}_\xi \right) &=0 \, . \label{eq:conservationII}
\end{align}
The fluxes are given by the following equations
\begin{align}
I^a &= \int_\Sigma d^4x \sqrt{-\gamma} J^a_{cons} \, , \\
H^a &= \int_\Sigma d^4x \sqrt{-\gamma} Q^a_{cons} \, . 
\end{align}
and, gauge fixing $A_r=0$, the rest of objects involved read
\begin{align}
J^a_{cons} &= F^{ra} + \frac{3}{2} J_A^a= F^{ra} + 2 \kappa \epsilon^{abcd} A_b F_{cd} \, , \\
k_\alpha^{ba} &= -F^{ba} + 2 \kappa \epsilon^{bacd} A_c \partial_r A_d \, , \\
Q^a_{cons} &= {t^a}_b \xi^b + \xi_c A^c \mathcal{J}^a \, , \\
\mathcal{J}^a &= F^{ra} +  J_A^a= F^{ra} + \frac{4}{3} \kappa \epsilon^{abcd} A_b F_{cd} \, , 
\end{align}
where ${t^a}_b$ is the Einstein-Hilbert Brown York tensor and the contributions to $k^{ba}_\xi$ only come from the current part, which can be easily confronted with the gauge formulas.

In order to radially integrate the equations we start by looking at the effects of the zero modes in \eqref{eq:conservationI} and \eqref{eq:conservationII}. First, we consider the $U(1)$ current. In this case, taking the $a$ index to be perpendicular to the Killing $\xi$, as is the magnetic field, one can show that the vorticity contributions cancel by the antisymmetry of the Levi-Civita tensor, while the only contribution that remains is the constant magnetic field times the radial derivative of $A_t$. Since the magnetic zero mode is constant we get the new conservation law
\begin{equation}
 \partial_r \int_\Sigma d^4 x \sqrt{-\gamma} \left( J_{cons}^a + 4 \kappa A_t \epsilon^{abcd} u_b \partial_c a_d \right)=0 \, .
\end{equation}
Integrating in the radial direction we get
\begin{equation}
\int d^4 x \sqrt{-\gamma_{(0)}} \langle J^a_{cons} \rangle= \int_H d^4 x \sqrt{-\gamma} J^a_{cons}(r_H) - 4 \kappa \mu B^a \, . 
\end{equation}
The boundary contribution due to the Bardeen-Zumino polynomials vanishes in our gauge and the horizon value of the conserved current almost trivially follows from the substitution of our ansatz into the Chern-Simons term
\begin{equation}
\int_{H} d^4 x \sqrt{-\gamma} J_{cons}^a ( r_H ) = - 4 \kappa \mu \left( B^a + 2 \mu \omega^a \right) \, .
\end{equation}
Which finally gives
\begin{equation}
\int d^4 x \sqrt{-\gamma_{(0)}} \langle J^a \rangle= - 8 \kappa \mu B^a  - 8 \kappa \mu^2 \omega^a \, .
\end{equation}

In the diffeomorphism case the treatment of the spatial derivative terms closely follows the previous one. The main difference is that, in this case, the presence of the $A_c \xi^c$ term will give an additional factor of $1/2$ due to the radial derivative of $A_t^2$. Then, the new conservation law is
\begin{equation}
\partial_r \int_{\Sigma} d^4 x \sqrt{-\gamma}\left( {t^a}_b \xi^b + A_c \xi^c \mathcal{J}^a  + \frac{4}{3}\kappa A_t^2 \epsilon^{abcd}u_b \partial_c a_d \right)=0 \, .
\end{equation}
In our gauge choice this gives the matching
\begin{equation}
\int d^4 x \sqrt{-\gamma_{(0)}} \langle {T^a}_b \xi^b \rangle = \int_H d^4 x \sqrt{-\gamma} A_c \xi^c \mathcal{J}^a + \frac{4}{3} \kappa \mu^2 B^a \, ,
\end{equation}
where we have already discarded the ideal contribution coming from the Brown York tensor at the horizon.

The evaluation of the consistent current is completely parallel to the covariant case, paying attention to the different coefficient. Finally we get
\begin{equation}
\int d^4 x \sqrt{-\gamma_{(0)}} \langle {T^a}_b \xi^b \rangle = - \int d^4 x \sqrt{-\gamma_{(0)}} \langle J^a_\epsilon \rangle= 4 \kappa \mu^2 B^a + \frac{16}{3} \kappa \mu^3 \omega^a \, .
\end{equation}

Although the coefficients coincide with those given in the introduction, we have no physical insight for the relationship between the further pieces coming from the continuity equation and the low energy physics of the anomalous transport phenomena.

\section{Conclusions}
We have extended the construction of membrane currents to anomalous theories. In doing so we identified how extrinsic contributions coming from gravitational Chern-Simons terms are linked to the thermal anomalous transport at the horizon. 
Such terms vanish at the conformal boundary but are dynamically generated at lower energies, finally giving the expected thermal effective action on the horizon. This is very reminescent of the Wilsonian integration of gapped excitations. Thus, it would be interesting to see if such a parallel can indeed be made and, in that case, how those modes have to be interpreted.

The horizon properties can be reformulated as CFT observables through the usage of conserved fluxes and we have shown that such fluxes coincide, up to subtleties in the $U(1)^3$ case, with \emph{conserved} membrane fields. This is a nontrivial extension of the previous arguments, allowing us to explain various results found in the literature in a simple and elegant way. 

Finally, holographic systems have long been used in the study of non-equilibrium processes and the first studies regarding the gravitational anomaly have been recently published \citep{Landsteiner:2017lwm}. It would be interesting to see if the membrane currents we have defined, which precisely account for these anomalous hydrodynamic fluctuations at the horizon, could be used to get analytical insight over such phenomena.

\section{Acknowledgements}
The authors would like to thank Karl Landsteiner for very useful discussions, participation in the early stages of the project and comments on early versions of the draft. This work is supported by FPA2015-65480-P and by the Spanish Research Agency (Agencia Estatal de Investigación) through the grant IFT Centro de Excelencia Severo Ochoa SEV-2016-0597. The work of J.F.-P. is supported by fellowship SEV-2012-0249-03. The work of C.C. is funded by Fundaci\'on La Caixa under ``La Caixa-Severo Ochoa'' international predoctoral grant.

\appendix

\section{Anomalies and currents} \label{appsubscurrentdef}

In this appendix we review some basic facts about anomalies in quantum field theory (for a review, see \cite{Bertlmann:1996xk}). In particular, we focus on the properties of consistent anomaly polynomials and the possible definition of current operators in the presence of external fields. As a concrete example we show how these are defined in the four dimensional case, so that they can be immediately related to the membrane currents of section \ref{sectionmembranecurrents}.

First, we define what we mean by a consistent anomaly. Let $W[A]$ be the generating functional of a quantum field theory with a classically conserved current $\mathcal{J}^a$ which couples to the external gauge field $A_a$. The conservation of the current is then reflected in the gauge invariance of the generating functional in the presence of external sources
\begin{equation}
D_a \mathcal{J}^a= 0 \, \leftrightarrow \, \delta_\alpha W[A]=0 \, .
\end{equation}
However, this is not necessarily true at the quantum level. In this case, one talks about a \emph{consistent} anomaly
\begin{equation}
\delta_\alpha W[A] =\int  \mathcal{A}_\alpha (A) \neq 0 \, .
\end{equation}
A key result of Wess and Zumino \citep{Wess:1971yu} is that the functionals $\mathcal{A}_\alpha$ are related to characteristic classes $P_{d+2}[A]$ in $d+2$ dimensions via a series of cohomological equations which are called the Wess-Zumino descent equations. The first two steps of these equations read
\begin{align}
P_{d+2}[A] &= d I_{CS}[A] \, , \\
\delta_\alpha I_{CS}[A] &= d \mathcal{A}_\alpha (A) \, ,
\end{align}
which connect the Chern-Simons actions in $d+1$ dimensions to the consistent anomalies in $d$ dimensions.

Let us now come to the properties of the current operators for anomalous theories. When no 't Hooft anomaly is present, the current is uniquely fixed by coupling the theory to an external gauge field $A_a$ for the global symmetry and differentiating the effective action with respect to it
\begin{equation}
\mathcal{J}^a = -i\frac{\delta}{\delta A_a} W[A,\gamma] \, . \label{AIconscurrent}
\end{equation}
The operator $\mathcal{J}^a$ defined in this way is gauge invariant and divergence-free.

In the anomalous case only one of the previous three properties can be imposed at once. This follows essentially from the Wess-Zumino procedure and from the definition of the consistent current $\mathcal{J}^a$, which makes it fail to be gauge invariant,
\begin{equation}
\delta_\alpha \mathcal{J}^a = -i\frac{\delta}{\delta A_a} \delta_\alpha W[A,\gamma]= -i \frac{\delta}{\delta A_a} \int \mathcal{A}_\alpha(A) \, .
\end{equation}
It is then natural to define these three different current operators:
\begin{itemize}

\item A \emph{consistent} current $\mathcal{J}^a$, defined through \eqref{AIconscurrent}, which fulfills the consistent anomaly equation
\begin{equation}
D_a \mathcal{J}^a =\mathcal{A} \, , \ \ \mathcal{A}=\frac{\delta }{\delta \alpha} W[A] \, .
\end{equation}
This current is not gauge invariant, as we have shown before, and it cannot be made so by any local counterterm.

\item A \emph{covariant} current $J^a$, which is invariant under gauge transformations
\begin{equation}
\delta_\alpha J^a=0\, .
\end{equation}
This can be obtained by adding a specific Bardeen-Zumino polynomial $J^a_{BZ}$ to the consistent current. It is made up of external fields to the conserved current and it fulfills
\begin{equation}
\delta_\alpha J_{BZ}^a= i \frac{\delta}{\delta A_a} \int \mathcal{A}_\alpha(A) \, . 
\end{equation}
As the Bardeen-Zumino polynomial is not divergence-free, the anomaly equation is changed
\begin{equation}
D_a J^a = \mathcal{A}_{cov}(A) \, ,
\end{equation}
where $\mathcal{A}_{cov}$ is a gauge invariant expression called the ``covariant'' anomaly.

\item A \emph{conserved} current $J^a_{cons}$ which is divergence-free
\begin{equation}
D_a J^a_{cons}=0 \, .
\end{equation}
However, it is neither gauge invariant nor related to the generating functional. By writing the anomaly as a total derivative, we can identify
\begin{equation}
J^a_{cons}= \mathcal{J}^a + \tilde{J}^a \, ,
\end{equation}
where the remaining piece is defined through
\begin{equation}
 D_a \tilde{J}^a = \mathcal{A} \, .
\end{equation}
In holography we saw that these conserved currents are the ones displaying nice RG properties for their long wavelength modes.
\end{itemize}

We will now review the examples relevant for our work. Let us start with the consistent $U(1)^3$ anomaly
\begin{equation}
D_a \mathcal{J}^a= - \frac{\kappa}{3} \epsilon^{abcd} F_{ab} F_{cd} \, .
\end{equation}
The gauge variation of the current is then
\begin{equation}
\delta_\alpha \mathcal{J}^a = -\frac{4}{3}\kappa \epsilon^{abcd} \partial_b \alpha F_{cd} \, ,
\end{equation}
which gives
\begin{equation}
J_{BZ}^a= \frac{4}{3}\kappa \epsilon^{abcd} A_b F_{cd} \, .
\end{equation}
The covariant anomaly is then
\begin{equation}
D_a J^a = -\kappa \epsilon^{abcd} F_{ab} F_{cd} \, .
\end{equation}
In order to obtain the conserved current, we start by rewriting the consistent anomaly
\begin{equation}
- \frac{\kappa}{3} \epsilon^{abcd} F_{ab} F_{cd}=- D_a\left( \frac{2}{3}\kappa \epsilon^{abcd} A_b F_{cd} \right) \, . 
\end{equation}
From here, it is clear that 
\begin{equation}
\tilde{J}_a= \frac{2}{3}\kappa \epsilon^{abcd} A_b F_{cd} = \frac{1}{2}J_{BZ}^a \, .
\end{equation}
These constructions can trivially be exported to the membrane currents in holography once the consistent current is defined from the bulk action.

For the consistent stress tensor $\mathcal{T}^{ab}$ the Ward identity reads
\begin{equation}
D_a \mathcal{T}^{ab} = F^{ba} \mathcal{J}_a - A^b D_a \mathcal{J}^a \, . \label{AIwitensor}
\end{equation}
In this case, the consistent stress tensor is also gauge invariant. Therefore, it will be equal to the covariant stress tensor $\mathcal{T}^{ab}=T^{ab}$. In fact, one can massage the Ward identity in the form
\begin{equation}
D_a T^{ab} = F^{ba}J_a \, ,
\end{equation}
by means of the identity
\begin{equation}
\xi_f A^f \epsilon^{abcd} F_{ab} F_{cd} = 4 \xi^f F_{fa} \epsilon^{abcd} A_b F_{cd} \, .
\end{equation}
In turn, looking at \eqref{AIwitensor} we can define a \emph{conserved} heat current
\begin{equation}
Q_{cons}^a= \mathcal{T}^{ab} \xi_b + A_b \xi^b \mathcal{J}^a \, .
\end{equation}
Notice, however, that this is not gauge invariant, in line with the general reasoning.

Next, in four spacetime dimensions we can have a mixed gauge-gravitational anomaly, coming from the anomaly polynomial
\begin{equation}
P_6[A, {\bf\Gamma}]= F P_2 [ {\bf R}] \, ,
\end{equation}
where $P_2$ is the second Pontriagyn class. In coordinates, $P_2[ {\bf{R}}] = \epsilon^{abcd} tr \left( {\bf{R}}_{ab} {\bf{R}}_{cd} \right)$. Being a mixed anomaly one can ``choose'' whether the consistent anomaly falls into the diffeomorphism or into the gauge sector. We start with a gauge invariant theory. Then the Bardeen counterterm
\begin{equation}
B[A,{\bf\Gamma}]= - 4 c \lambda \int d^4 x \sqrt{-\gamma} \epsilon^{abcd} A_a tr \left( {\bf \Gamma}_b \partial_c {\bf \Gamma}_d + \frac{2}{3} {\bf \Gamma}_b {\bf \Gamma}_c {\bf \Gamma}_d \right) 
\end{equation}
gives the consistent Ward identities
\begin{align}
D_a \mathcal{J}^a &= c \lambda \epsilon^{abcd} tr \left( {\bf{R}}_{ab} {\bf{R}}_{cd} \right) \, , \\
D_a \mathcal{T}^{ab} - F^{ba} \mathcal{J}_a + A^b D_a \mathcal{J}^a &= (1-c) 2\lambda \gamma^{bc} \frac{1}{\sqrt{-\gamma}}\partial_a \left(\sqrt{-\gamma} \epsilon^{defg}F_{de}\partial_f\Gamma_{gc}^a  \right) \, . \label{eq:AIWI}
\end{align} 
The current is gauge invariant, but not diffemorphism invariant due to the gravitational anomaly, while the stress tensor is neither gauge nor diffeomorphism invariant. The simplest way to construct the covariant stress tensor is to notice that, for $c=1$, the only nontrivial variation is 
\begin{equation}
\delta_\alpha \mathcal{T}^{ab}= -4 \lambda \epsilon^{mnp(a} D_e\left( D_m \alpha {R_{np}}^{b) e} \right) \, ,
\end{equation}
which gives 
\begin{equation}
T^{ab}= \mathcal{T}^{ab} + 4 \lambda  \epsilon^{mnp(a} D_e\left( A_m {R_{np}}^{b) e} \right)\, .
\end{equation}
The further gauge dependence for $c\neq 1$ comes from the Bardeen-Zumino counterterm. It gives
\begin{align}
J^a &= \mathcal{J}^a - (1-c) \lambda \epsilon^{abcd} tr \left( {\bf \Gamma}_b \partial_c {\bf \Gamma}_d + \frac{2}{3} {\bf \Gamma}_b {\bf \Gamma}_c {\bf \Gamma}_d \right)= \mathcal{J}^a -(1-c) J^a_{{\bf \Gamma}} \, , \\
T^{ab}&= \mathcal{T}^{ab} + 4 c \lambda  \epsilon^{mnp(a} D_e\left( A_m {R_{np}}^{b) e} \right) + (1-c) \frac{1}{2} D_d \left( l^{dba} + l^{adb} - l^{abd} + (a \leftrightarrow b) \right)  \, ,
\end{align}
where ${l^{ab}}_c$ is given by 
\begin{equation}
 {l^{ab}}_c = 2 \lambda \epsilon^{aefg} F_{ef} \Gamma_{gc}^b \, .
\end{equation}
It can be seen to coincide with the definition \eqref{eq:labc}. For the conserved currents one gets, from
\begin{equation}
c \lambda \epsilon^{abcd} tr\left( {\bf R}_{ab}{\bf R}_{cd} \right)= 4 D_a \left(c \lambda \epsilon^{abcd} tr \left({\bf \Gamma}_b \partial_c {\bf \Gamma}_d + \frac{2}{3} {\bf \Gamma}_b {\bf \Gamma}_c {\bf \Gamma}_d \right) \right) \, , 
\end{equation}
that
\begin{equation}
J^a_{cons}= \mathcal{J}^a - c J^a_{{\bf \Gamma}} \, .
\end{equation}
The construction of the conserved heat current $Q^a$ is more tricky if a diffeomorphism anomaly is present. In order to succeed in such construction it is important to remember that the presence of a heat current is tied with the presence of a Killing vector in the background spacetime. Things can get problematic if an anomaly is present in the diffeomorphism sector, as in this case the general covariance is broken and some additional conditions have to be imposed on the background connection in order for the external fields to be invariant under the action of the Killing field. In particular, we would like the diffeomorphism variation of the connection to vanish
\begin{equation}
0=\delta_\xi {\bf \Gamma}= \mathcal{L}_\xi {\bf\Gamma} + d {\bf \Lambda} \, .
\end{equation}
It turns out that, in order to connect to the holographic results in the main text, it is useful to impose this condition in a noncovariant way, as in \citep{Bonora:2011gz}. Thus we choose a coordinate system in which $d {\bf \Lambda} =0$ and, separately, $\mathcal{L}_\xi {\bf \Gamma}=0$.

Once this is imposed, the conserved heat current can indeed be constructed by repeated integration by parts starting from the diffeomorphism Ward identity, discarding terms proportional to $d {\bf \Lambda}$. Contracting then \eqref{eq:AIWI} with $\xi_b$ and noticing that
\begin{equation}
\begin{aligned}
(1-c) 2 \lambda \xi^b \partial_a \left(\sqrt{-\gamma} \epsilon^{defg}F_{de}\partial_f\Gamma_{gb}^a  \right)&= \partial_a \left( 2(1-c) \lambda \sqrt{-\gamma} \epsilon^{defg} F_{de} \partial_f \Gamma_{gb}^a \xi^b \right)- \\
& - \partial_a \left( 2(1-c) \lambda \sqrt{-\gamma} \epsilon^{adef} F_{de} \Gamma_{f b}^g \Lambda^b_g \right) + O(\partial \Lambda) \, ,
 \end{aligned}
\end{equation}
we can use the Killing condition on the rest of \eqref{eq:AIWI} to recover a conservation law
\begin{equation}
\frac{1}{\sqrt{-\gamma}}\partial_a \left( \sqrt{-\gamma} Q^a_{cons} \right)=0 \, ,
\end{equation}
where
\begin{equation}
Q^a_{cons} = {\mathcal{T}^a}_b \xi^b + A_c \xi^c \mathcal{J}^a - (1-c) \frac{1}{\sqrt{-\gamma}} \partial_b \left( \sqrt{-\gamma} {l^{ba}}_c \xi^c \right)  + 2 ( 1 - c) {l^{(ab)}}_c \Lambda_b^c \, . \label{qftheatcurrent}
\end{equation}

\section{Presymplectic current for Chern-Simons theories}\label{AIIpresCS}
In this appendix we briefly derive the general form of the presymplectic current $\theta$ for Chern-Simons terms. This will be used in sections 3 and 4 to derive the general form of the constraint equations and the Komar charge.

In order to derive it, let us vary the general Chern-Simons action as
\begin{equation}
\delta I_{CS} = \left( \frac{\partial I_{CS}}{\partial F} \delta F+  \frac{\partial I_{CS}}{\partial A} \delta A\right) + \left( \frac{\partial I_{CS}}{\partial d {\bf \Gamma}} d\delta {\bf \Gamma}  +  \frac{\partial I_{CS}}{\partial {\bf \Gamma}} \delta {\bf \Gamma} \right) \, . 
\end{equation}
Integrating by parts once leads to
\begin{equation}
\delta I_{CS} = \Sigma^\mu \delta A_\mu + E^{\mu\nu\rho} \delta \Gamma_{\mu\nu}^\rho + d \left( \frac{\partial I_{CS}}{\partial F} \delta A + \frac{\partial I_{CS}}{\partial d {\bf \Gamma}} \delta {\bf \Gamma} \right) \, .
\end{equation}
Finally, using the explicit expression for the Christoffel symbols in terms of the metric one gets the final answer
\begin{equation}
\delta I_{CS} = \Sigma^\mu \delta A_\mu - \nabla_\rho \Sigma^{\mu\nu\rho} \delta g_{\mu\nu} + \nabla_\mu \left( 2 \frac{\partial I_{CS}}{\partial F_{\mu\nu}} \delta A_\nu + \frac{\partial I_{CS}}{\partial \partial _\mu {\bf \Gamma}_\rho} \delta {\bf \Gamma}_\rho + \Sigma^{\alpha\beta \mu} \delta g _{\alpha \beta}  \right) \, ,
\end{equation}
from which one reads the presymplectic current
\begin{equation}
\theta^\mu = 2 \frac{\partial I_{CS}}{\partial F_{\mu\nu}} \delta A_\nu + 2 \frac{\partial I_{CS}}{\partial  {\bf R}_{\mu\rho}} \delta {\bf \Gamma}_\rho + \Sigma^{\alpha\beta \mu} \delta g _{\alpha \beta}   \, . \label{presymplecticcurrent}
\end{equation}
From this formula one can readily derive the constraint equations and the bulk expressions for the membrane current as, following Wald
\begin{equation}
\int_{\Sigma} d^d x \sqrt{-g} n_\mu \left(\theta^\mu_{\xi} - \Xi^{\mu}_\xi \right) \doteq \int_{\Sigma} d^d x \sqrt{-\gamma}\left(C^a \xi_a + D_a Q^a_\xi\right) \, ,
\end{equation}
where $\xi$ is an r-independent vector field tangent to $\Sigma$, $C^a$ are the diffeomorphism constraints and $Q^a$ is the Wald charge. The same reasoning is of course valid when the diffeomorphism is substituted be an $r$-independent gauge transformation.

\section{Thermal Hall effect and topological massive gravity}\label{appthermalhall}
A similar story as the one we have told in the main body also applies to the simplest case with gravitational anomalies, which is topological  massive gravity. In this appendix we put forward some interesting connection between the presence of the higher derivative term $u^{ab}$, the Cauchy problem and the development of thermal transport at the horizon.

As we have already anticipated the theory is defined by Einstein gravity with cosmological constant together with the Chern Simons three form
\begin{equation}
I_{CS} = {\bf \Gamma} d {\bf \Gamma} + \frac{2}{3} \bf \Gamma^3 \, ,
\end{equation}
and is a prototype for the dual description of the 1+1 dimensional gravitational anomaly
\begin{equation}
\delta_\Lambda I_{CS} = \int_{\Sigma} \Lambda d \Gamma \, ,
\end{equation} 
which gives the consistent anomaly \citep{Skenderis:2009nt}
\begin{equation}
A_a =  \lambda \epsilon^{mn}\partial_b\partial_m \hat{\Gamma}_{na}^{b} \, .
\end{equation}
This has to be contrasted with the known result for the consistent gravitational anomaly for a $CFT$ with left and right central charges $c_L, c_R$, which gives
\begin{equation}
\lambda = \frac{c_L - c_R}{96 \pi^2} \, .
\end{equation}

Such chiral theories are known to display a thermal Hall effect related to the presence of the anomaly, which basically comes from the Schwarzian transformation of the stress tensor
\begin{equation}
T'(z')= \left(\frac{d z'}{d z}\right)^{-2}\left(T(z) -\frac{c_L}{12} Schw(z',z)\right) \, .
\end{equation}
For a thermal state this leads to an energy current
\begin{equation}
J_\epsilon = \frac{T^2}{24} \left( c_L - c_R\right) \, .
\end{equation}
From the point of view of holography, the analysis of this theory is somewhat subtle due to the Cauchy problem at the conformal boundary. In fact, it was shown in \citep{Skenderis:2009nt} that the most general asymptotic boundary conditions allow for two independent modes in the metric expansion
\begin{equation}
\gamma_{ab} \sim e^{2r} \left( r b_{ab} + \hat{\gamma}_{ab} + \dots \right) \, ,
\end{equation}
leading to a situation in which two degenerate operators of dimension two, $t^{ab}$ and $s^{ab}$, can be turned on at the boundary. They correspond with the two free data of the solution
\begin{equation}
t^{ab}= \frac{2}{\sqrt{-\gamma}} \frac{\delta S_g}{\delta \hat{\gamma}_{ab}} \, , \ \ s^{ab}= \frac{2}{\sqrt{-\gamma}} \frac{\delta S_g}{\delta b_{ab}}\, .
\end{equation}
Their correlators are those of a logarithmic CFT and thus violate unitarity. One can however recover a unitary theory by suppressing the $b_{ab}$ mode in the asymptotic expansion, which is the strategy we have followed in the main text.
One has to stress, however, that such condition does not imply that this dynamical mode is suppressed throughout the bulk. Indeed this suggests a comparison with the on-shell variation \eqref{eq:onshellvariationanomalies}, where two independent modes are associated with the metric and the extrinsic curvature respectively.

Indeed, given that the construction of the various quantities follows closely the five dimensional case by getting rid of the field strengths, one can check that the logarithmic mode is related to the non-vanishing of a boundary observable as
\begin{equation}
\lim_{r \to + \infty} \sqrt{-\gamma} u^{ab} = 8 \lambda \left( \hat{\epsilon}^{ab} {b_m}^b + \hat{\epsilon}^{bm} {b_m}^a  \right) \, .
\end{equation}
where we have expanded linearly in $b_{ab}$. This, together with the conservation equations for the membrane stress tensor, gives a suggestive picture. In a thermal state, even if asymptotically AdS boundary conditions are imposed, the mode conjugate to $b_{ab}$ is dynamically generated along the RG flow. On the stretched horizon, it generates the thermal Hall effect
\begin{equation}
\lim_{r \to r_H} \sqrt{-\gamma} {\Theta^i}_b \xi^b = \lim_{r \to r_H} 2 \sqrt{-\gamma} u^{ic} K_{cb} \xi^b = 4 \lambda T^2 \, ,
\end{equation}
which matches once the correct value is substituted for $\lambda$.

In light of this admittedly hand-waving similarity, it would be interesting if a more precise link could be given by studying the anomalous flow to low energies of a thermal chiral CFT.

\bibliography{AnomTrans}{}
\bibliographystyle{JHEP}

\end{document}